\documentclass[12pt,preprint]{aastex}

\newcommand{\be}{\begin{equation}}
\newcommand{\ee}{\end{equation}} 
\newcommand{\simless}{\lower.5ex\hbox{$\; \buildrel < \over \sim\;$}}
\newcommand{\simgreat}{\lower.5ex\hbox{$\; \buildrel > \over \sim\;$}} 

\shorttitle{Cosmic-ray Energy Diffusion}
\shortauthors{Fatuzzo \& Melia}

\begin{document}

\title{A Numerical Assessment of Cosmic-ray Energy Diffusion through Turbulent Media}

\author{M. Fatuzzo}
\affil{Physics Department, Xavier University, Cincinnati, OH 45207}
\email{fatuzzo@xavier.edu}

\author{F. Melia}
\affil{Department of Physics, The Applied Math Program, and Steward Observatory, \\ 
The University of Arizona, AZ 85721}
\email{fmelia@email.arizona.edu}

\begin{abstract}
How and where cosmic rays are produced, and how they diffuse through various
turbulent media, represent fundamental problems in astrophysics with far reaching 
implications, both in terms of our theoretical understanding of high-energy processes 
in the Milky Way and beyond, and the successful interpretation of space-based and 
ground based GeV and TeV observations. For example, recent and ongoing detections, 
e.g., by Fermi (in space) and HESS (in Namibia), of $\gamma$-rays produced in regions 
of dense molecular gas hold important clues for both processes. In this paper, we carry 
out a comprehensive numerical investigation of relativistic particle acceleration and
transport through turbulent magnetized environments in order to derive broadly 
useful scaling laws for the energy diffusion coefficients.
\end{abstract}


\section{Introduction}

Until recently, the widely held paradigm for the origin of cosmic rays, at least
below the knee at roughly $1$ PeV, promoted the view that all of the intragalactic
injection occurred via first-order Fermi acceleration in supernova shells. But
any direct evidence for this view is meager and equivocal. More recently, data from
balloon-borne experiments have refuted the expectation from supernova
acceleration schemes that the cosmic-ray spectrum ought to be structureless
and universal (Wefel 1988). The latest measurements with PAMELA
confirm and extend the balloon-based claims (Adriani et al. 2011). These data
seem to call for a diverse variety of acceleration sites and mechanisms throughout
the Galaxy.  

In recent work (Fatuzzo \& Melia 2011, 2012a, 2012b), we have begun to assess
the feasibility of stochastic acceleration within turbulent magnetized regions using
highly detailed simulations of individual particle trajectories. A principal goal
of this work is to accurately determine the spatial and energy diffusion coefficients
of cosmic-ray protons in a broad range of environments, e.g., inside molecular 
clouds and the more tenuous intercloud medium. The spatial and energy diffusion 
coefficients calculated over a broad range of parameter space may be used, e.g., 
to compare estimates of the time required to energize protons up to TeV energies 
with the escape and cooling times throughout the interstellar medium. In previous
applications, this approach has allowed us to conclude that protons in the intercloud 
medium at the galacitc center can be energized up to the $1-10$ TeV energies 
required to account for the observed HESS emission in this region (Aharonian et al.
2006; Liu et al. 2006a; Ballantyne et al. 2007; Wommer et al. 2008; see also
Markoff et al. 1997, 1999; Liu \& Melia 2001).  Stochastic particle acceleration 
by magnetic turbulence appears to be a viable mechanism for cosmic-ray production 
at the galactic center (Liu et al. 2006b).  

Developing an understanding of how the spatial diffusion coefficients depend
on the physical environment and energy of the particles was the subject of our
previous work (Fatuzzo et al. 2010, 2011). The focus of the present paper is specifically 
to determine the diffusion of cosmic rays in energy space, with its attendant dependence on all of
the critical physical characteristics of the medium, such as the magnetic intensity,
degree of turbulence, and size of the fluctuations. As before, we do this by using
a modified numerically based formalism developed for the general study of cosmic-ray 
diffusion by Giacalone \& Jokipii (1994). This approach has already been used 
successfully in several other contexts (see, e.g., O'Sullivan et al 2009; Fatuzzo
\& Melia 2012a, 2012b). Here, we will extend this robust numerically-based framework 
for the general analysis of stochastic acceleration of cosmic-rays by the 
turbulent electric fields generated along with the time-dependent turbulent 
magnetic field in a dynamically active medium.

We stress that our adopted model of turbulence as a superposition of Alfv\'enic like waves 
with linear dispersion relations is not meant to fully describe MHD turbulence in the interstellar
medium (ISM).  For example, the model does not account for the fact that magnetic fluctuations decorrelate 
due to non-linear interactions before they can propagate over distances of multiple wavelengths
(Goldreich \& Sridhar 1995) -- an effect that 
leads to resonance broadening and as such, influences how thermal particles interact with turbulence
(Lynn et al. 2012; 2013).   However, our intent is to adopt a formalism that adequately 
describes turbulence as seen ÒlocallyÓ by highly relativistic cosmic rays.  Since relativistic particles 
have speeds that are much greater than the Alfv\'en speeds considered in our analysis 
(which is then the limit in which our results are expected to be valid), they should not be 
sensitive to dynamical processes that occur on MHD timescales.   In essence, we are relying 
on basic principles (e.g., the scaling laws between intensity and wavelength) 
to capture the global features of MHD turbulence that affect the propagation 
and acceleration of high energy cosmic rays.  

The work presented here extends the results of previous works, most notably, 
that of O'Sullivan et al. (2012), as it significantly broadens the explored parameter space 
and also considers anisotropically distributed wave vectors (Goldreich \& Sridhar 1995; Cho \& Vishniac 2000).
We focus primarily on strong turbulence ($\delta B \sim B$) for which quasilinear theory is not
applicable, although we do present results for isotropic weak turbulence as a consistency check.  
Our paper is organized as follows. The pertinent properties of the medium 
through which the particles diffuse are outlined in \S 2. The scheme for 
generating the turbulent magnetic and electric fields is presented in \S 3, 
along with the equations that govern the motion of cosmic rays. The
basic elements of stochastic acceleration are discussed in \S4, and the results
of our work are presented in \S 5. The conclusions of our work are 
summarized in \S 6.

\section{The Physical Medium}
The physical parameters found throughout the intergalactic and interstellar
media have values that span several orders of magnitude.  Of particular
interest to our study, the most vacuous regions of the intergalactic medium
have particle number densities $n \simless 10^{-3}$ cm$^{-3}$ and magnetic 
field strengths $B \simless 0.1 \mu$G (see, e.g., Kronberg 1994; Fraschetti 
\& Melia 2008). In contrast, the denser regions near the supermassive black 
hole at the galactic center have densities $n \ga 10^{12}$ cm$^{-3}$ and field strengths 
$B \ga 1$ G (Ruffert \& Melia 1994; Falcke \& Melia 1997; Kowalenko \& Melia
1999; Melia 2007; see also Misra \& Melia 1993 for the case of stellar-mass 
size black holes).  

Exactly how the magnetic field is partitioned within these various media is 
not yet known, but there is a general relation between density and field strength. 
In the simplest case where flux freezing applies (say in the ISM), the magnetic 
field strength $B$ would scale with gas density $n$ according to $B\propto n^{1/2}$.
It is noteworthy, then, that an analysis of magnetic field strengths measured 
in molecular clouds yields a relation between $B$ and $n$ of the form
\begin{equation}
B \sim 10 \,\mu\hbox{\rm G} \left({n \over 10^2\, \hbox{\rm cm}^{-3}}
\right)^{0.47}\;,
\end{equation}
though with a significant amount of scatter in the data used to produce this fit 
(Crutcher 1999; see also Fatuzzo et al. 2006 and references cited therein). 
This result is consistent with the idea that nonthermal linewidths, 
measured to be $\sim$$1$ km s$^{-1}$ throughout the cloud environment 
(e.g., Lada et al. 1991), arise from MHD fluctuations.  

Of course, even among molecular clouds, the physical environment can be
quite different depending on location. Near the galactic center, the molecular 
clouds are considerably different from their counterparts in the disk. For example, 
the average molecular hydrogen number density over the Sgr B complex---the 
largest molecular cloud complex (Lis \& Goldsmith 1989; Lis \& Goldsmith 1990; 
Paglione et al. 1998) near the galactic center---has a density $3$--$10 \times 
10^3$ cm$^{-3}$. Like its traditional counterparts, Sgr B displays a highly nonlinear 
structure, containing two bright sub-regions, Sgr B1 and Sgr B2, the latter having
an average molecular density of $\sim 10^6$ cm$^{-3}$, and containing three 
dense ($n \sim 10^{7.3-8}$ cm$^{-3}$), small ($r\sim 0.1$ pc) cores---labeled 
North, Main and South. These cores also show considerable structure, containing 
numerous ultra-compact and hyper-compact HII regions.  As such, the densities 
associated with the galactic-center molecular clouds are about two orders of 
magnitude greater than those in the disk of our galaxy.

The exact nature of magnetic turbulence in these environments is itself not well
constrained, though magnetic fluctuations typically have a power-law spectrum.
Their intensity at a given wavelength scales according to $(\delta B_\lambda)^2 \sim 
\lambda^{\Gamma-1}$, with values of $\Gamma = 1$ (Bohm), $\Gamma = 3/2$ (Kraichnan) 
or $\Gamma = 5/3$ (Kolmogorov) often adopted. In addition, the range in wavelengths 
over which these fluctuations occurs is not well known, though it is reasonable to 
assume that the upper end corresponds to the lengthscale over which the fluctuations 
are generated. (For example, in the ISM, the turbulence is generated by supernova
remnants and stellar-wind collisions, so one might expect the longest wavelength 
to be on the order of several parsecs or less; see, e.g., Coker \& Melia 1997;
Melia \& Coker 1999.) The lower end must be smaller than the characteristic length 
scale associated with the particle motion (i.e., the gyration radius), under the 
assumption that magnetic energy ultimately dissipates into plasma energy via its 
coupling to these particles.

\section{Governing Equations}
We explore how cosmic rays diffuse through a homogeneous hydrogen gas of mass 
density $\rho = n m_p$ threaded by a uniform static background field ${\bf B_0}$, on 
which magnetic and electric fluctuations propagate. From our numerical simulations,
we then compute energy diffusion coefficients covering a broad range of particle 
energies and parameter space expected to span the great diversity of environments 
observed both within our galaxy and in the intergalactic medium. We focus primarily 
on strong turbulence, for which the energy density of the turbulent fields is comparable 
to that of the background fields. 

A standard numerical approach to studying cosmic-ray diffusion 
treats  the spatially fluctuating
magnetic component  $\delta{\bf B}$ as the superposition of a large number
of  randomly polarized  transverse waves with wavelengths $\lambda_n 
= 2\pi/k_n$, logarithmically spaced between $\lambda_{min}$ and $\lambda_{max}$
(e.g., Giancoli \& Jokipii 1994; Casse et al. 2002;  Fatuzzo 
et al. 2010). Adopting a static turbulent field removes the necessity of specifying 
a dispersion relation between the wavevectors $k_n$ and their corresponding angular 
frequencies $\omega_n$.  This approach appears suitable for considering
highly non-linear turbulence ($\delta B >> B_0$), or simply an environment without
a background component. Of course, turbulent magnetic fields in cosmic 
environments are not static. Nevertheless, a static formalism in spatial 
diffusion calculations of relativistic particles seems justified for environments 
in which the Alfv\'en speed is much smaller than the speed of light.
 
The situation in this paper is quite different: we are focusing on the energy 
diffusion of cosmic rays propagating through a turbulent magnetic field, which 
requires the use of a time-dependent formalism in order to self-consistently include 
the fluctuating electric fields that must also be present (say, from Faraday's law). 
At present, such a theory of MHD turbulence in the interstellar medium remains
elusive. Nevertheless, it is generally understood that turbulence is driven from a cascade of
longer wavelengths to shorter wavelengths as a result of wave-wave interactions. For strong
MHD turbulence in a uniform medium, this cascade seemingly produces eddies on small
spatial scales that are elongated in the direction of the underlying magnetic  field, so that the
components of the wave vector along ($k_{||}$) and across ($k_\perp$) the underlying field direction 
are related by the expression $k_{||}\propto k_\perp^{2/3}$, with a Kolmogorov 
energy spectrum that scales as $k_\perp^{-5/3}$ (Goldreich \& Sridhar 1995;
Cho \& Lazarian 2003).

It is beyond the scope of this paper to develop a self-consistent theory of MHD turbulence 
in the ISM  that includes
electric field fluctuation.  We therefore adopt the formalism of O'Sullivan et al (2012).
Specifically, we assume a medium represented by a nonviscous, perfectly conducting fluid 
threaded by a uniform static field ${\bf B_0} $, and use linear MHD 
theory as a guide. In the linear regime, one can encounter three types of MHD 
waves: Alfv\'en, fast and slow.  As in O'Sullivan et al. (2009), we here consider
only Alfv\'en waves, for which the turbulent magnetic field may be written as
a sum of $N$ randomly directed waves
\be
 \delta {\bf B} = \sum_{n=1}^N {\bf A_n} \,  e^{i( {\bf k_n} \cdot {\bf r}-
\omega_n t+\beta_n)}\,.
\ee
As noted already, this formalism does not adequately describe decorrelation effects
known to be important for thermal cosmic rays (Lynn et al. 2012; 2013).  
But it should serve as a reasonable model for the turbulence experienced by particles
with velocities much greater than the Alfv\'en speed $v_A$, as such particles would be expected to 
travel over many correlation lengths ($\sim 0.1 \lambda_{max}$) in an Alfv\'en time
$\tau_A \sim \lambda_{max} / v_A$.

To keep the analysis as broad as possible, we focus on 
an isotropic turbulence spectrum, but we also perform a suite
of experiments with anisotropic turbulence as informed by the results
of Golreich \& Sridhar (1995).  For the isotropic case,
the direction of each propagation vector ${\bf k_n}$ is set through a 
random choice of polar angles $\theta_n$ and $\phi_n$, and the phase of 
each term is set through a random choice of $\beta_n$.  
The sum has $N = N_k \log_{10}[k_{max}/k_{min}]$ terms, with $k_n$ evenly 
spaced on a logarithmic scale between $k_{min} = 2\pi / \lambda_{max}$ and
$k_{max} = 2\pi/\lambda_{min}$.  A value of $N_k$ = 25 appears to provide 
enough terms to suitably model what is really a continuous rather than a discrete 
system. The appropriate choice of $\Gamma$ in the scaling
\begin{equation}
A_n^2 = A_1 ^2\left[{k_n \over k_1} \right]^{-\Gamma}
{\Delta k_n\over \Delta k_1} 
= A_1^2\left[{k_n \over k_1} \right]^{-\Gamma+1}
\end{equation}
sets the desired spectrum of magnetic turbulence (e.g., $\Gamma = 3/2$ 
for Kraichnan and 5/3 for Kolmogorov turbulence). For our adopted scheme, 
the value of $\Delta k_n / k_n$ is the same for all $n$. The normalization
for $A_1$ is set by the parameter
\be
\eta = {\langle \delta B^2 \rangle \over B_0^2}\;,
\ee
defined as the ratio of magnetic 
energy density in the turbulent component to that of the static background 
field $B_0$. 

For the anisotropic case,  the direction of each perpendicular 
wavevector $k_{n\perp}$ is set through a 
random choice of azimuthal angle $\phi_n$, and the phase of 
each term is again set through a random choice of $\beta_n$.  
The corresponding parallel component of the wavevector is then set through the
relation
\be
k_{n||} = \pm {\sqrt{2}\over 2} k_{1\perp}^{1/3}\,k_{n\perp}^{2/3}\,
\ee
(with a randomly chosen sign),
such that
\be
k_1 = \sqrt{k_{1\perp}^2+k_{1||}^2} = {2\pi \over \lambda_{max}}\;.
\ee
Consistent with the results of Goldreich \& Sridhar (1995), we only consider
a Kolmogorov profile, so that
\begin{equation}
A_n^2 
= A_1^2\left[{k_{n\perp} \over k_{1\perp}} \right]^{-2/3}\,.
\end{equation}

Since Alfv\'en waves don't compress the fluid through which they propagate, 
their fluid velocity ${\bf v}$ satisfies the condition ${\bf k} \cdot {\bf v} = 0$. In addition,
${\bf v} \cdot {\bf B_0} = 0$ for Alfv\'en waves. The fluid velocity 
associated with the $n$-th term in Equation (2) is therefore
\be
\delta {\bf v_n} = \pm A_n \, {v_A\over B_0} \, {{\bf B_0} \times {\bf k_n} 
\over | {\bf B_0} \times {\bf k_n} |}
 \,  e^{i( {\bf k_n} \cdot {\bf r}-\omega_n t+\beta_n)}\,,
\ee
where the sign is chosen randomly for each term in the sum.  The dispersion
relation for Alfve\'nic waves is given by the expression  
\be
\omega_n = v_A |k_{n\|}|\,.
\ee
Each wave has a magnetic field given by the linear form of 
Amp\`ere's Law,  
\be
{\bf  A_n} = \mp A_n \, {{\bf k_n} \cdot {\bf B_0} \over | {\bf k_n} \cdot 
{\bf B_0} |} \, {{\bf B_0} \times {\bf k_n} \over | {\bf B_0} \times {\bf k_n} |}
 \,,
 \ee
which is identical to that of O'Sullivan et al. (2009).   

Insofar as the electric fields are concerned, if we were to naively extrapolate
from the results of linear MHD theory, the total electric field $\delta {\bf E}$ 
associated with the turbulent magnetic field in Equation~(2) would be given
by a sum over the terms
\be
 \delta{\bf E_n}=  -\delta {\bf v_n} \times {\bf B_0}\,.
\ee
Notice that $ \delta{\bf E} \cdot  {\bf B_0}$ = 0, but the second order 
term $\delta {\bf E}\cdot \delta {\bf B} \ne 0$.  The presence of an electric 
field component parallel to the magnetic field in this second order term can 
significantly increase the acceleration efficiency artificially, especially if the 
formalism is extended to the nonlinear regime ($\delta B \sim$$B_0$).  
However, the interstellar medium is highly conductive, so any electric field 
component parallel to the magnetic field should be quickly quenched.  
O'Sullivan et al. (2009) circumvented this problem by first obtaining the 
total fluid velocity $\delta {\bf v}$ via the summation
\be
\delta {\bf v} = \sum_{n=1}^N  \delta {\bf v_n}\,,
\ee
and then using the MHD condition to set the total electric field:
\be
\delta {\bf E} = - {\delta {\bf v}\over c} \times{\bf B}\;,
\ee
where ${\bf B} = {\bf B_0} + \delta{\bf B}$. This is the procedure we
too will use here.

\begin{figure}[h]
\figurenum{1}
\begin{center}
{\centerline{\epsscale{1.0} \plotone{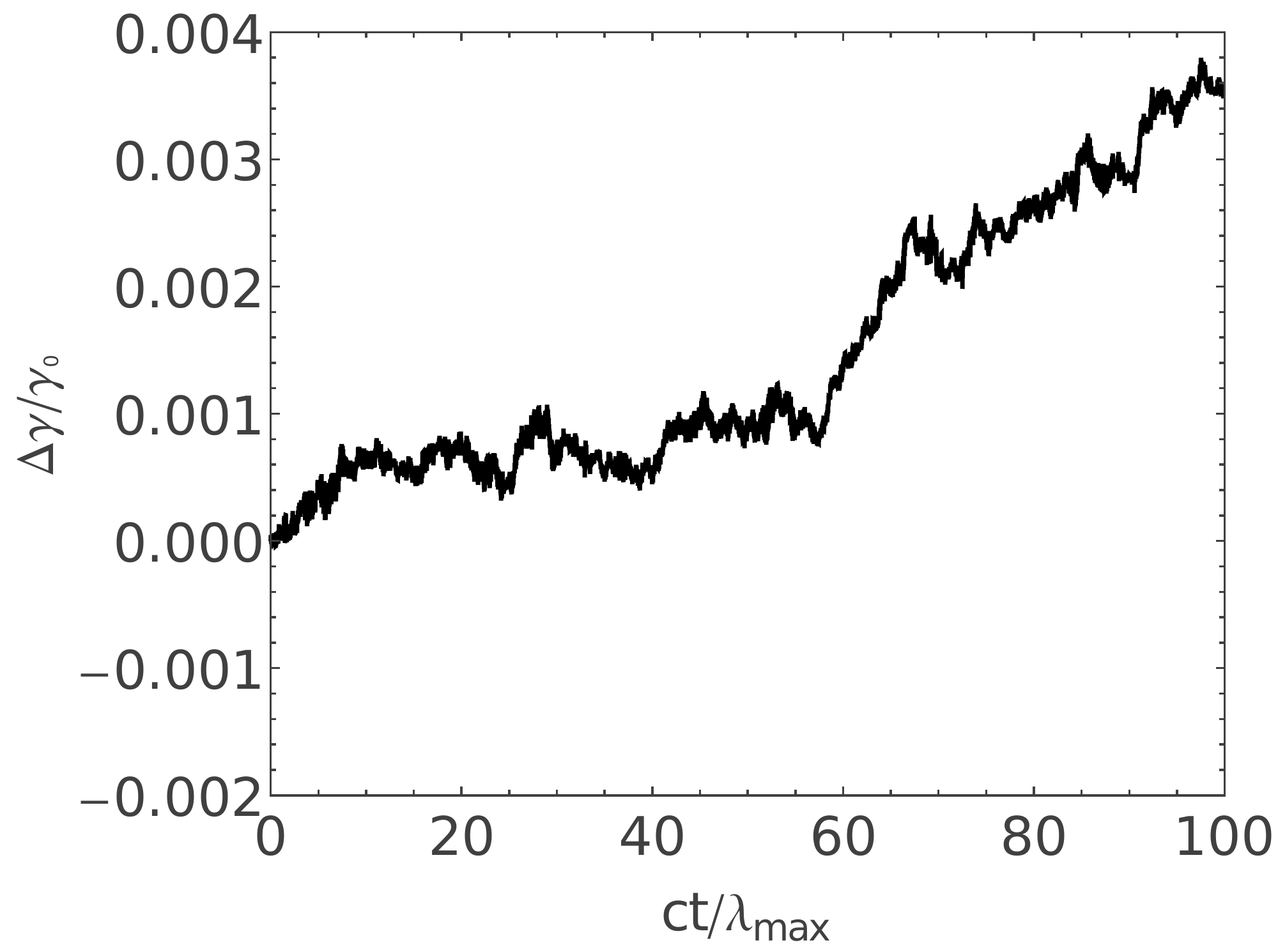} }}
\end{center}
\vskip-0.3in
\figcaption{The fractional change in particle energy
$\Delta\gamma / \gamma_0$
as a function of time for a $\gamma_0 = 10^5$ particle
injected into an environment  
 ($B_0 = 100\,\mu$G, $n = 10^2$ cm$^{-3}$)
 with an isotropic Alfv\'enic turbulent field defined by the parameters
$\lambda_{max} = 0.1$ pc, $\Gamma = 5/3$, and $\eta = 1.0$.  
}
\end{figure}

With these electric and magnetic field components, one may then solve
the Lorentz force equation
\be
{d\over  dt}  (\gamma m_p {\bf v})=  e \left[\delta {\bf E} +
{{\bf v} \times {\bf B}\over c}\right]\;,
\ee
with
\be
{d {\bf r}\over dt} = \bf v\;,
\ee
to determine the motion of a relativistic charged proton with Lorentz
factor $\gamma$ through the turbulent medium. The solutions to these 
equations  are not sensitive to the value of $\lambda_{min}$ so long as 
the particle's gyration radius $R_g = \gamma mc^2/(eB) >> \lambda_{min}$ 
(Fatuzzo et al. 2010).  As such, we set $\lambda_{min} = 0.1 \gamma mc^2/(eB_0)$ 
in all our simulations. To completely specify the physical state of an environment
to be studied, we must therefore provide values for the parameters $B_0$, $n$, 
$\lambda_{max}$, $\Gamma$ and $\eta$.

\section{Basic Elements of Stochastic Acceleration}
The motion of a charged particle in perpendicular uniform 
electric and magnetic fields represents a fundamental topic
in electrodynamics and is well understood.  It is therefore easy to 
show that in general, the charge gains and loses kinetic 
energy in cyclic fashion as it
``drifts" in the ${\bf E} \times {\bf B}$ direction.  

\begin{figure}[h]
\figurenum{2}
\begin{center}
{\centerline{\hskip-0.15in\epsscale{1.0} \plotone{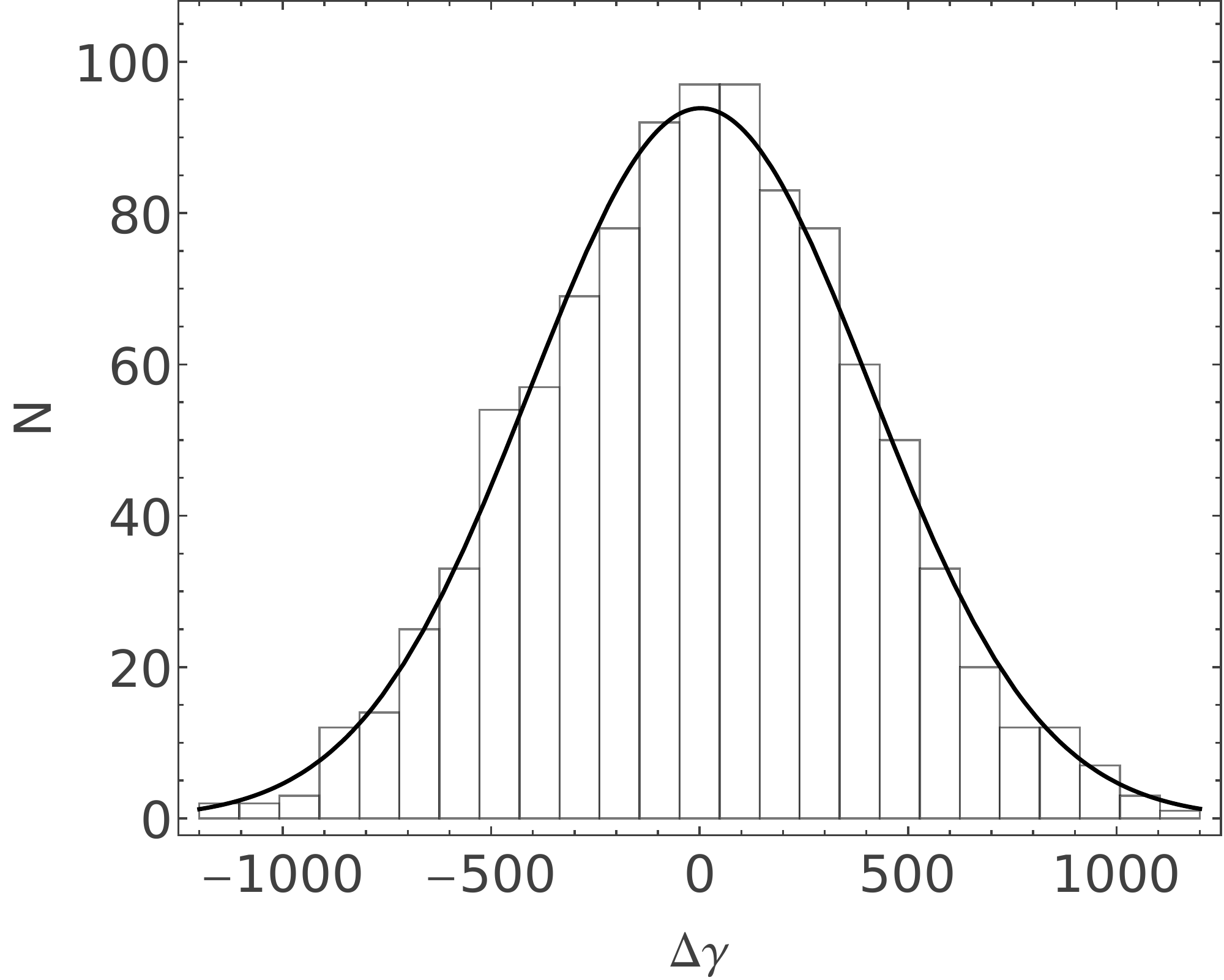} }}
\end{center}
\vskip-0.3in
\figcaption{The particle energy distribution at time $t = 1000 \lambda_{max}/c$ 
for an ensemble of
$N_p = 1,000$ particles injected into an environment  
 ($B_0 = 100\,\mu$G, $n = 10^2$ cm$^{-3}$)
 with an anisotropic Alfv\'enic turbulent field defined by the parameters
$\lambda_{max} = 0.1$ pc, $\Gamma = 5/3$, and $\eta = 1.0$.  
The solid line shows a Gaussian fit to the data.
}
\end{figure}

\begin{figure}[h]
\figurenum{3}
\begin{center}
{\centerline{\hskip-0.15in\epsscale{1.0} \plotone{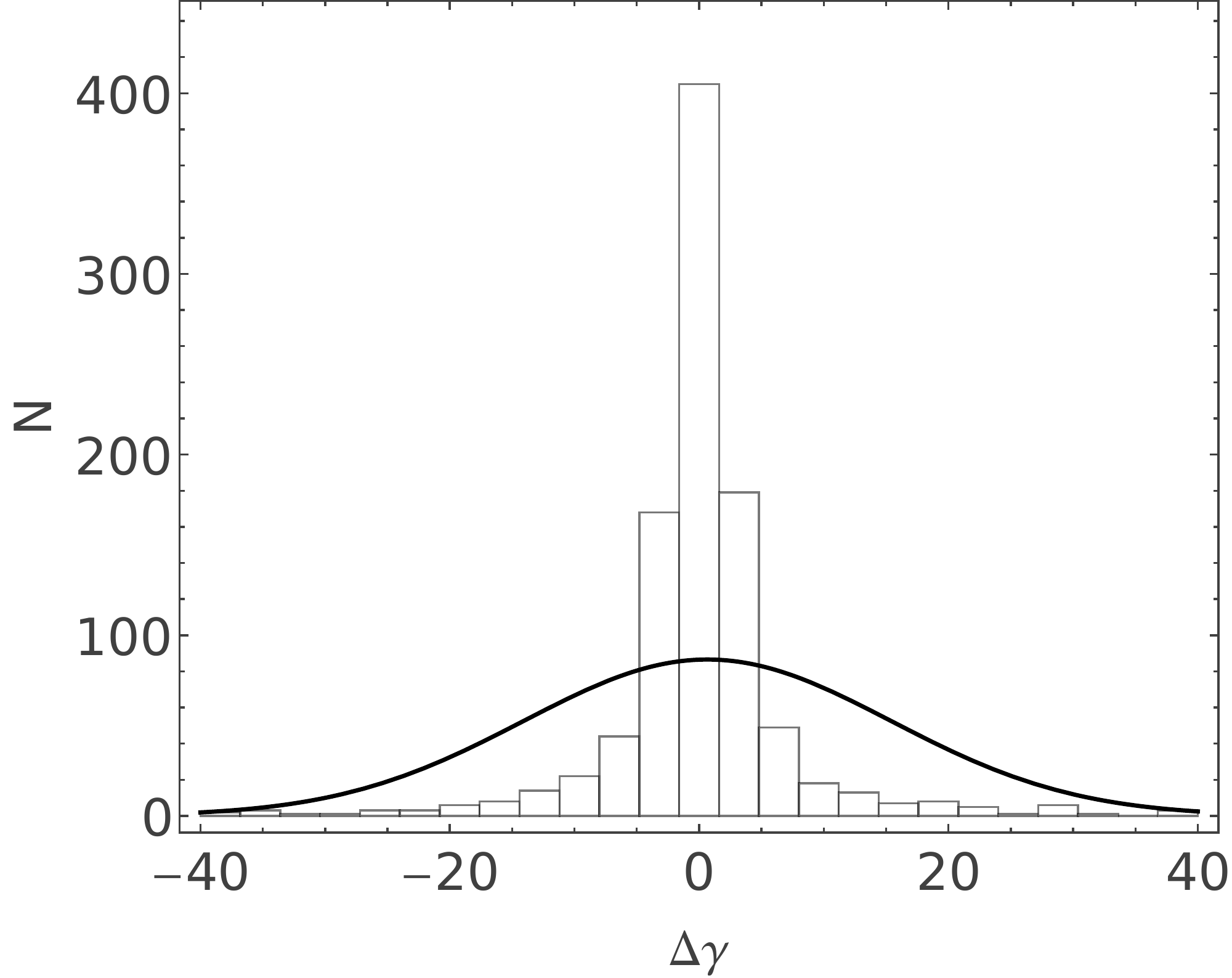} }}
\end{center}
\vskip-0.3in
\figcaption{The particle energy distribution at time $t = 1000 \lambda_{max}/c$ 
for the ensemble of
$N_p = 1,000$ particles injected into an environment  
 ($B_0 = 100\,\mu$G, $n = 10^2$ cm$^{-3}$)
 with an anisotropic Alfv\'enic turbulent field defined by the parameters
$\lambda_{max} = 0.1$ pc, $\Gamma = 5/3$, and $\eta = 0.01$.  
The solid line shows a Gaussian fit to the data.
}
\end{figure}

In the turbulent medium in which the magnetic and electric fields are 
perpendicular, a single particle's energy, characterized by $\Delta\gamma / \gamma_0 = \gamma / \gamma_0 - 1$,  
therefore exhibits a random-walk like behavior, as shown in figure~1. The energy distribution 
of an ensemble of particles injected with the same Lorentz factor $\gamma_0$ thus broadens 
as the particles sample the turbulent nature of the accelerating 
electric fields.  Interestingly, these distributions appear Gaussian for all of the isotropic
and $\eta = 1$ anisotropic cases explored in our analysis.  This point is illustrated by figure~2, 
which shows the distribution of $\Delta\gamma$ values at time $t = 1000 \lambda_{max}/c$
for an ensemble of $1,000$ particles injected with $\gamma_0 = 10^5$ into 
a medium defined by the same parameters used to calculate the
energy evolution shown in figure~1, but with an anisotropic turbulence profile.  
In such cases, one can  
therefore quantify the stochastic acceleration 
of particles in turbulent fields through the dispersion of the resulting
distributions of initially mono-energetic particles.  

In contrast, the particle distributions obtained for anisotropic, weak turbulence ($\eta \ll 1$) 
cases have wings that are much broader than a Gaussian distribution, so that
their dispersion significantly overestimates the true distribution width.  This point
is illustrated in figure 3, which shows the  distribution of $\Delta\gamma$ values 
at time $t = 1000 \lambda_{max}/c$
for an ensemble of $1,000$ particles injected with $\gamma_0 = 10^5$ into 
a medium defined by the same parameters used to calculate the
distribution shown in figure~2, but with $\eta = 0.01$.  

To further  compare strong and weak isotropic and anisotropic 
turbulence, we 
plot in figure~4 the dispersion $\sigma_\gamma\equiv\sqrt{\langle\Delta\gamma^2\rangle}$ of the
$\Delta\gamma$ distribution as a 
function of time for the particles used to generate figures~2 and 3
(for which the turbulence was anisotropic) along with their isotropic counterparts.
As found throughout our analysis, there is little difference between the isotropic and
anisotropic cases when the turbulence is strong ($\eta =1$).  However, anisotropic turbulence
appears to be significantly less effective at energizing particles than isotropic turbulence when
$\eta \ll 1$, in agreement with results obtained in the quasi-linear approximation
(Chandran 2000;  see also Yan \& Lazarian 2002).

\begin{figure}[h]
\figurenum{4}
\vskip 0.1in
\begin{center}
{\centerline{\hskip-0.15in\epsscale{1.0} \plotone{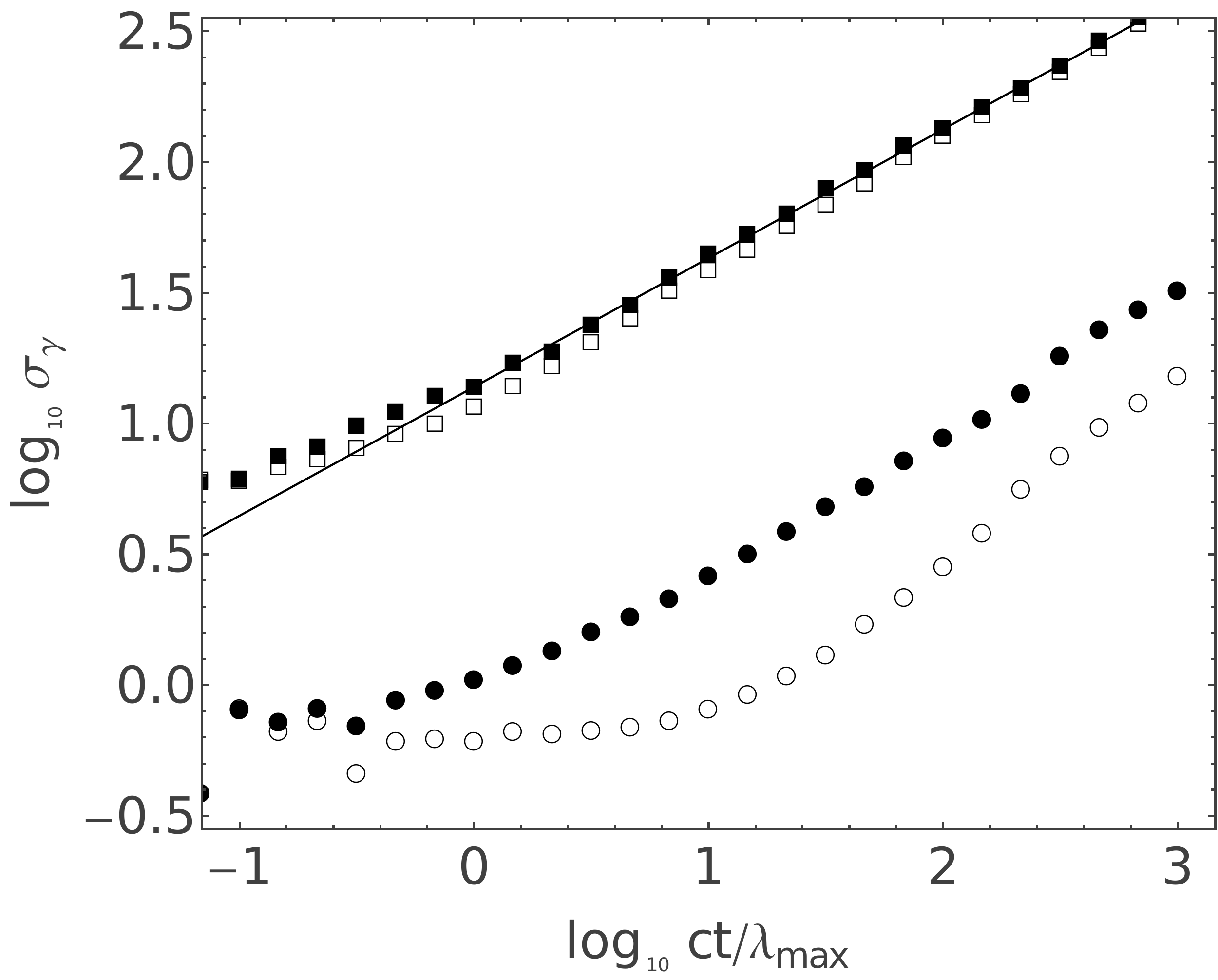} }}
\end{center}
\vskip-0.3in
\figcaption{The dispersion of the particle energy distribution as a function 
of time for the $N_p = 1,000$ particles injected into an environment  
 ($B_0 = 100\,\mu$G, $n = 10^2$ cm$^{-3}$)
 with an Alfv\'enic turbulent field defined by the parameters
$\lambda_{max} = 0.1$ pc, and  $\Gamma = 5/3$.  Solid squares:  isotropic 
turbulence with $\eta = 1.0$. Open squares: anisotropic turbulence with 
$\eta = 1.0$.  Solid circles:  isotropic 
turbulence with $\eta = 0.01$. Open circles: anisotropic turbulence with 
$\eta = 0.01$. 
The solid line serves as a reference and has a slope of $1/2$, clearly
indicating that $\sigma_\gamma\propto\sqrt{t}$ for time 
$t \ga \lambda_{max}/c$.
}
\end{figure}

Clearly, the chaotic nature of motion through 
turbulent fields necessitates a statistical analysis. We define a single experiment 
as a numerical investigation of particle dynamics through a given environment (as 
defined by the parameters $B_0$, $n$, $\Gamma$, $\lambda_{max}$, and $\eta$)
over a broad range of particle injection energies (as defined by the Lorentz factor $\gamma_0$).
As expected from the random nature of stochastic
acceleration, $\sigma_\gamma \propto \sqrt{t}$ once particles have had
a chance to sample the turbulent nature of the underlying electric fields, i.e., for
$t \ga \lambda_{max}/c$.  This in turn means that the energy diffusion coefficient
$D_\gamma\equiv \langle \Delta \gamma^2 \rangle / (2 \Delta t)$ 
can be calculated by using an integration time $\Delta t \ga \lambda_{max}/c$.
For each particle energy, we numerically integrate the equations of motion for a time
$\Delta t = 10 \lambda_{max}/c$ for $N_p = 1,000$ protons randomly injected  
from the origin. Each particle samples its own unique magnetic field structure
(i.e., the values of $\beta_n$, $\theta_n$, $\phi_n$ and the choice of a $\pm$ are 
chosen randomly for each particle for the isotropic case).

\section{Results of Numerical Experiments}
We use the procedure described above to carry out a suite of
experiments sampling a broad region of parameter space
for isotropic turbulence, and perform a limited complimentary set
of experiments for anisotropic, strong ($\eta = 1$) Kolmogorov turbulence. 
We limit our analysis
to particle energies for which the gyration radius $R_g$ falls comfortably 
below the maximum turbulence wavelength $\lambda_{max}$, so that
particles actually diffuse through the turbulent medium.  
A principal goal of this paper is to determine the relationship between 
the energy diffusion coefficient $D_\gamma$ and the particle's energy
for each experiment.  We therefore fit the numerical data
using a power law 
\be
D_\gamma = D_{\gamma 0} \gamma^{\alpha}\,.
\ee
The parameters for each experiment and
corresponding values of $D_{\gamma 0}$ and $\alpha$ obtained
through the fits are summarized in Table 1.  As a reference, we then
compare these empirical fits to the quasi-linear expression
\be
D^{ql}_\gamma \approx {v_A^2\over c^2} \left({\delta B\over B_0}\right)^2
\left({R_g\over\lambda_{max}}\right)^{\Gamma-1} {\gamma^2 c\over R_g} \,,
\ee
(Schlickeiser 1989, O'Sullivan et al. 2009).  We note that in terms of our parameters, this
predicted expression reduces to the simpler form
\be
D^{ql}_\gamma\propto  n^{-1} B_0^{4-\Gamma} \eta \,\lambda_{max}^{1-\Gamma}\, \gamma^\Gamma\,.
\ee

\begin{deluxetable}{rccccccc}
\tablecolumns{8}
\tablewidth{0pc}
\tablecaption{Summary of Experiments}
\tablehead{
\colhead{Exp} & \colhead{$B_0$ ($\mu$G)}   & \colhead{$n$ (cm$^{-3}$)}    
& \colhead{$\Gamma$} &
\colhead{$\lambda_{max}$ (pc)}  &  \colhead{$\eta$}&  \colhead{$D_{\gamma 0}$ (s$^{-1}$)}
&  \colhead{$\alpha$}}
\startdata
1 & 100  & 100 & 5/3 & 0.1 & 1&$5.4\times 10^{-13}$ &1.44 \\
2 & 100  & 100 & 5/3 & 0.1 & 0.1&$2.5\times 10^{-14}$ &1.48 \\
3 & 100  & 100 & 5/3 & 0.1 & 0.01&$2.6\times 10^{-16}$ &1.63 \\
4 & 100  & 100 & 5/3 & 0.1 & 0.001&$5.9\times 10^{-18}$ &1.70 \\
5 & 100  & 100 & 3/2 & 0.1 & 1&$ 1.2\times 10^{-12} $ &1.39 \\
6 & 100  & 100 & 3/2 & 0.1 & 0.001&$2.4\times 10^{-17}    $ &1.64 \\
7 & 100  & 100 & 1 & 0.1 & 1&$6.1\times 10^{-11} $ &1.10 \\
8 & 100  & 100 & 1 & 0.1 & 0.001&$8.2\times 10^{-15}  $ &1.20 \\
9 & 100  & 1 & 5/3 & 0.1 & 1&$4.7\times 10^{-11}$ &1.46 \\
10 & 100  & 3.16 & 5/3 & 0.1 & 1&$1.5\times 10^{-11}$ &1.46 \\
11 & 100  & 10 & 5/3 & 0.1 & 1&$3.9\times 10^{-12}$ &1.47 \\
12 & 100  & 31.6 & 5/3 & 0.1 & 1&$1.4\times 10^{-12}$ &1.46 \\
13 & 100  & 316 & 5/3 & 0.1 & 1&$1.2\times 10^{-13}$ &1.48 \\
14 & 100  & $10^3$ & 5/3 & 0.1 & 1&$5.2\times 10^{-14}$ &1.45 \\
15 & 0.1  & 100 & 5/3 & 0.1 & 1&$1.1\times 10^{-20}$ &1.46 \\
16 & 3.16  & 100 & 5/3 & 0.1 & 1&$7.5\times 10^{-17}$ &1.45 \\
17 & 3160  & 100 & 5/3 & 0.1 & 1&$3.0\times 10^{-9}$ &1.46 \\
18 & $10^5$  & 100 & 5/3 & 0.1 & 1&$4.0\times 10^{-6}$ &1.54 \\
19 & 100  & 100 & 5/3 & $10^{-4}$ & 1&$1.0\times 10^{-11}$ &1.48 \\
20 & 100  & 100 & 5/3 & 100 & 1&$1.8\times 10^{-14}$ &1.47 \\
21 & 0.1  & 100 & 3/2 & 0.1 & 1&$1.9\times 10^{-20}$ &1.39 \\
22 & 3.16  & 100 & 3/2 & 0.1 & 1&$1.6\times 10^{-16}$ &1.37 \\
23 & 3160  & 100 & 3/2 & 0.1 & 1&$1.4\times 10^{-8}$ &1.37 \\
24 & $10^5$  & 100 & 3/2 & 0.1 & 1&$5.3\times 10^{-5}$ &1.42 \\
25 & 100  & 100 & 3/2 & $10^{-4}$ & 1&$1.8\times 10^{-11}$ &1.40 \\
26 & 100  & 100 & 3/2 & 100 & 1&$1.0\times 10^{-13}$ &1.38 \\
27 & 0.1  & 100 & 1 & 0.1 & 1&$1.1\times 10^{-19}$ &1.11 \\
28 & 3.16  & 100 & 1 & 0.1 & 1&$2.6\times 10^{-15}$ &1.09 \\
29 & 3160  & 100 & 1 & 0.1 & 1&$9.6\times 10^{-7}$ &1.12 \\
30 & $10^5$  & 100 & 1 & 0.1 & 1&$6.2\times 10^{-2}$ &1.07 \\
31 & 100  & 100 & 1 &$10^{-4}$ & 1&$1.0\times 10^{-10}$ &1.12 \\
32 & 100  & 100 & 1 & 100 & 1&$2.6\times 10^{-11}$ &1.10\\
\enddata
\end{deluxetable}

\begin{figure}[h]
\figurenum{5}
\begin{center}
{\centerline{\hskip-0.15in\epsscale{1.0} \plotone{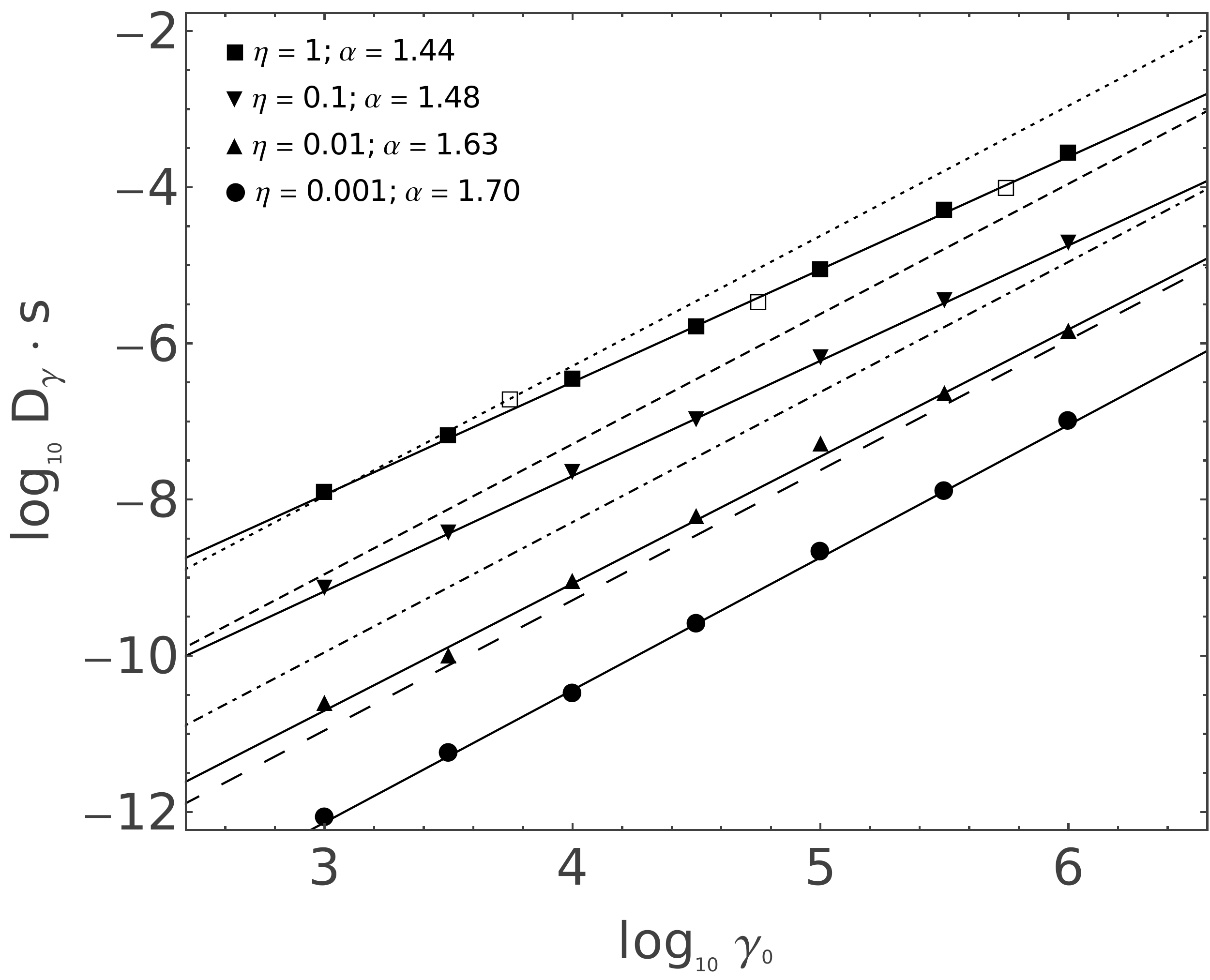} }}
\end{center}
\vskip -0.3in
\figcaption{Energy diffusion coefficients as a function of particle
Lorentz factor for Experiments 1--4 which sample various turbulence 
strengths, $\eta$. Solid shapes denote values for isotropic turbulence,
and open squares denote values for anisotropic turbulence.  The uniform
magnetic field is $100\;\mu$G in every case. In addition, $n=100$
cm$^{-3}$, and $\lambda_{max}=0.1$ pc.  Solid lines show fits 
to the isotropic turbulence data.  Results obtained by 
quasi-linear theory as given by Eq. (17) are shown by the
dotted ($\eta = 1$), dashed ($\eta = 0.1$), dot-dashed ($\eta = 0.01$)
and long-dashed ($\eta = 0.001$) lines.  
}
\end{figure}

We plot the results of Experiments 1--4 in Figure 5 and 
Experiments 1, 4, and 5--8 in Figures 6 and 7 to illustrate how
the diffusion coefficient index changes between the strong 
($\eta \sim 1$) and weak ($\eta \ll1$) turbulence limits.  
As predicted by quasi-linear theory, 
$\alpha \approx \Gamma$ when $\eta \ll1$ for isotropic turbulence, with the best 
match occurring for Kolomogorov diffusion.  However, 
Equation (17) overestimates the value of $D_{\gamma 0}$
by as much as two orders of magnitude. 
In addition, the indices decrease as the turbulence grows stronger,
and quasi-linear theory does not appear to adequately 
describe the energy dependence for Kolmogorov ($\Gamma = 5/3$) and
Kraichnan ($\Gamma = 3/2$) diffusion when $\eta \sim 1$.

\begin{figure}[h]
\figurenum{6}
\begin{center}
{\centerline{\hskip-0.15in\epsscale{1} \plotone{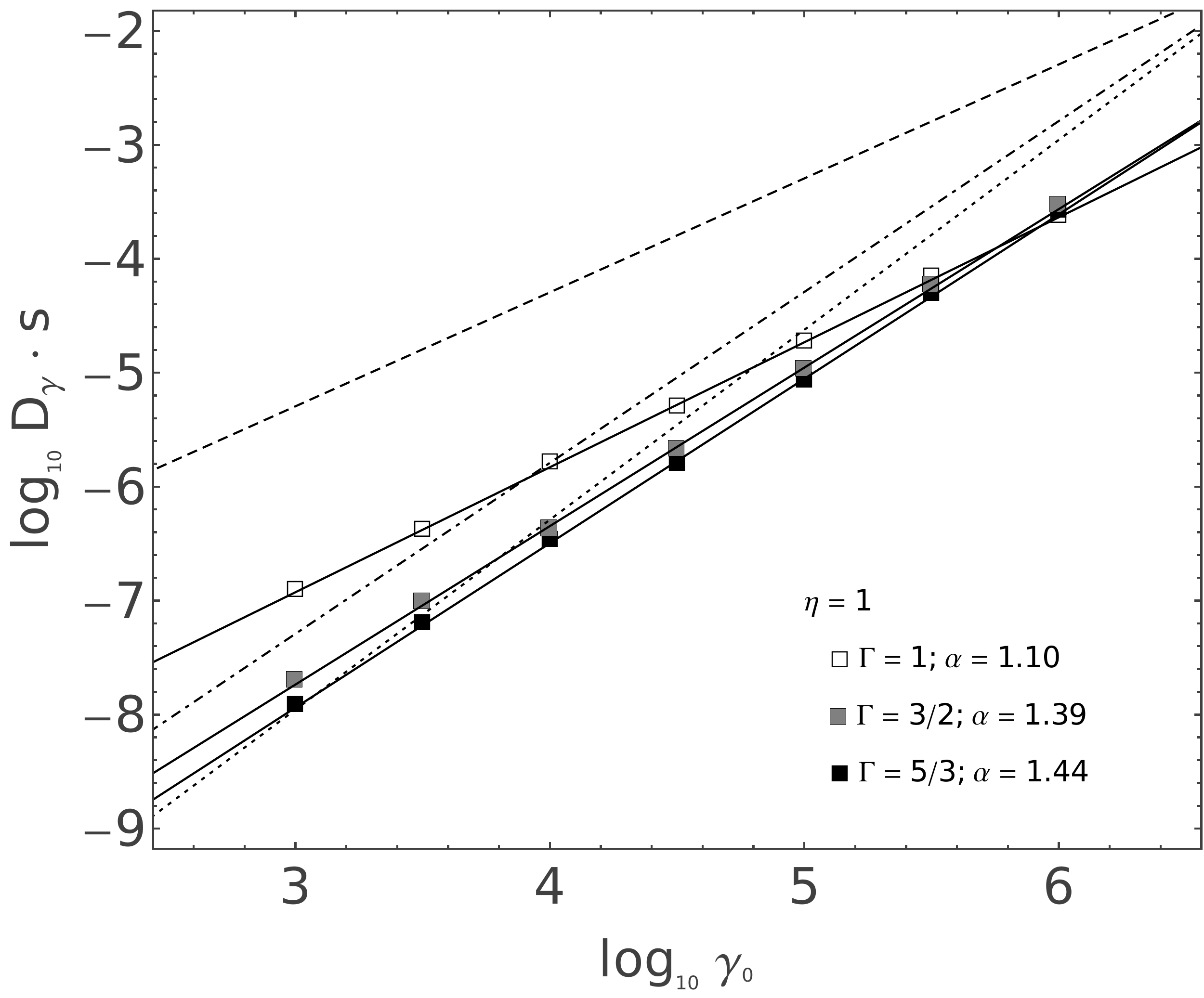} }}
\end{center}
\vskip -0.3in
\figcaption{Energy diffusion coefficients as a function of particle
Lorentz factor for Experiments 1, 5, and 7 (isotropic turbulence). The uniform
magnetic field is $100\;\mu$G in every case. In addition, $n=100$
cm$^{-3}$, and $\lambda_{max}=0.1$ pc.  Solid lines show fits 
to the data. Results obtained by 
quasi-linear theory as given by Eq. (17) are shown by the
dotted ($\Gamma = 5/3$), dot-dashed ($\Gamma = 3/2$),  and dashed ($\Gamma = 1$) 
lines.  }
\end{figure}

\begin{figure}[h]
\figurenum{7}
\begin{center}
{\centerline{\hskip-0.15in\epsscale{1} \plotone{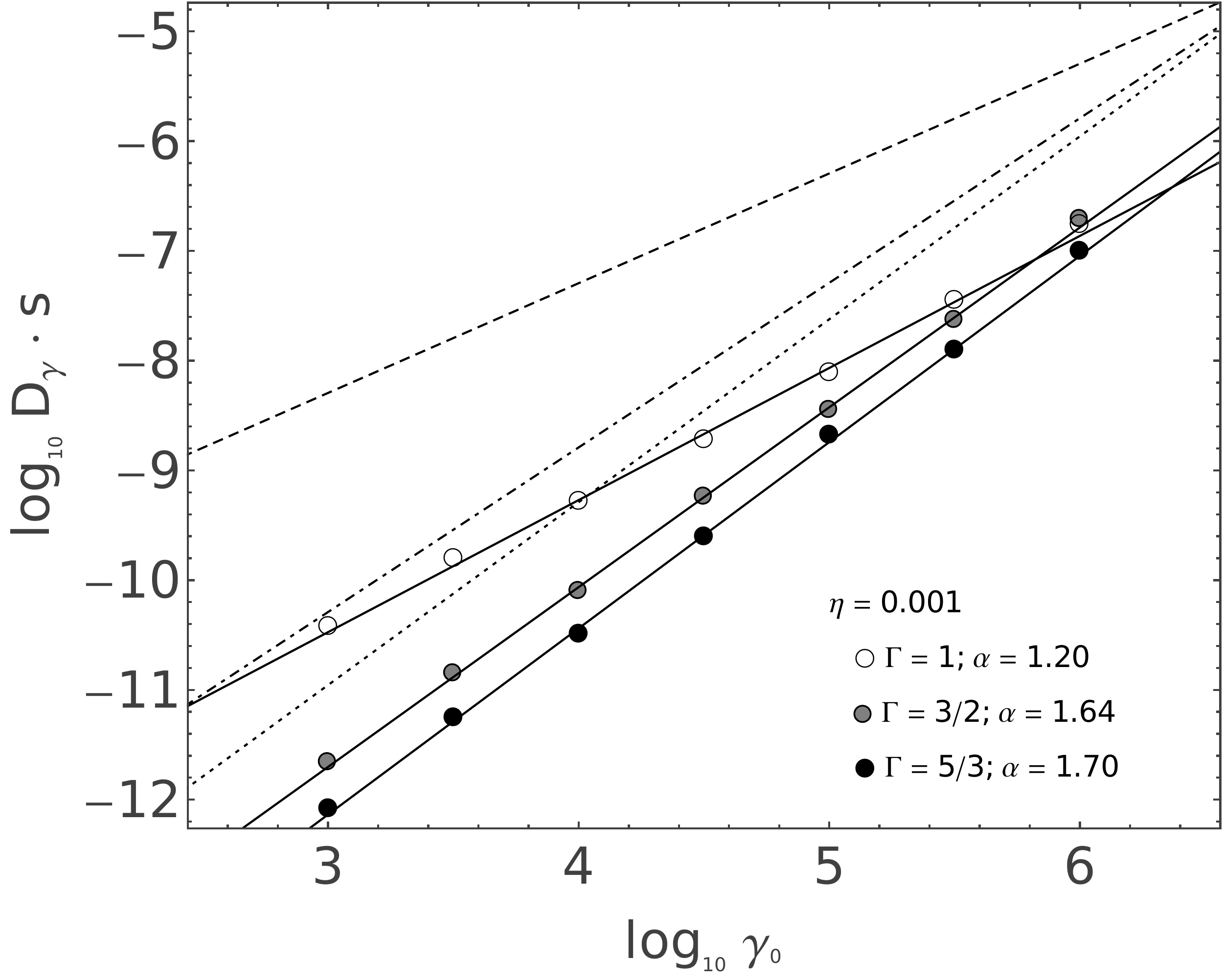} }}
\end{center}
\vskip -0.3in
\figcaption{Energy diffusion coefficients as a function of particle
Lorentz factor for Experiments 4, 6 and 8 (isotropic turbulence). The uniform
magnetic field is $100\;\mu$G in every case. In addition, $n=100$
cm$^{-3}$, and $\lambda_{max}=0.1$ pc.  
Solid lines show fits 
to the data. Results obtained by 
quasi-linear theory as given by Eq. (17) are shown by the
dotted ($\Gamma = 5/3$), dot-dashed ($\Gamma = 3/2$),  and dashed ($\Gamma = 1$) 
lines.  }
\end{figure}

We next consider how the diffusion coefficients for 
Kolmogorov ($\Gamma = 5/3$), Kraichnan ($\Gamma = 3/2$) and
Bohm ($\Gamma = 1$) turbulence scale with $n$, $B_0$ and 
$\lambda_{max}$, but limit our analysis to the strong turbulence limit
($\eta = 1$).  Our results for Kolmogorov turbulence (derived
from experiments 1 and 9--20) are shown in Figures 8--13.   
Specifically, Figure 8 illustrates the energy dependence of
$D_\gamma$ for four values of ambient density $n$. As expected,
a greater density yields a smaller Alfv\'en speed, and
therefore a reduced energy diffusion.  But the index $\alpha$ 
is not sensitive to $n$.  We note that Equation (17) overestimates 
the energy diffusion coefficient for $\gamma_0 \ga 10^3$,
and underestimates the energy diffusion coefficient for $\gamma_0 \la 10^3$.  
Figure 9 then illustrates the relation between $D_\gamma$ and $n$ for the value
$\gamma_0 = 10^5$, where the fits from Figure 8 (along with fits to the data not shown
if that figure) are used to obtain
the plotted data points.  The relationship between $D_\gamma$ and $n$
clearly takes on the same scaling obtained from quasi-linear theory, 
i.e., $D_\gamma \propto n^{-1}$.  This result is expected since $D_\gamma \propto$
$(\Delta\gamma)^2\propto$ $(\delta E)^2 \propto$ $v_A^2 \propto n^{-1}$.  We therefore
assume that this scaling holds for all values of $\Gamma$.
\begin{figure}[h]
\figurenum{8}
\begin{center}
{\centerline{\hskip-0.15in\epsscale{1.0} \plotone{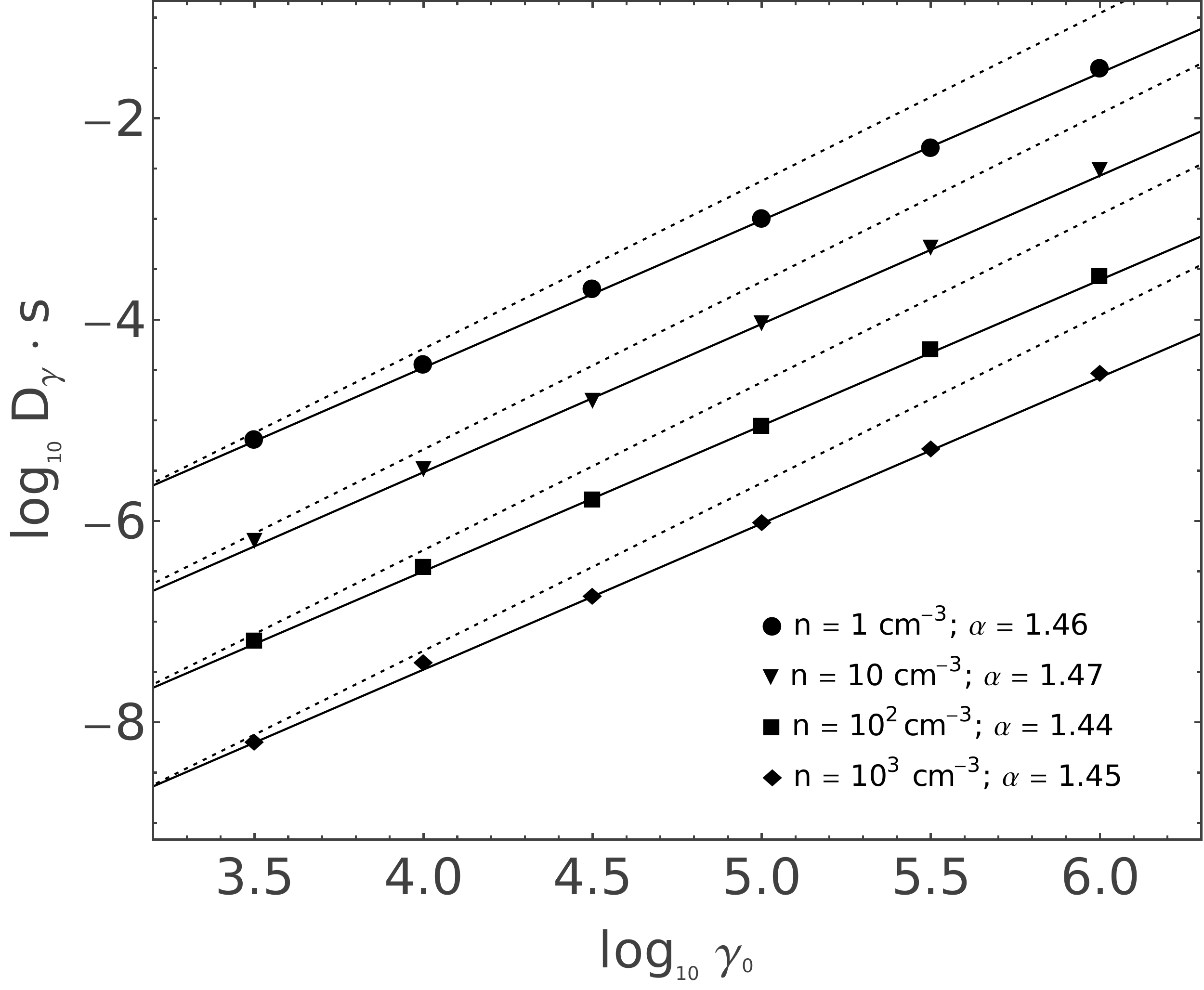} }}
\end{center}
\vskip -0.3in
\figcaption{Energy diffusion coefficients as a function of particle
Lorentz factor for Experiments 1,9, 11 and 14 (isotropic turbulence), which sample various 
ambient densities $n$. The uniform
magnetic field is $100\;\mu$G,
$\eta = 1$, $\Gamma=5/3$, and $\lambda_{max}=0.1$ in every case. 
Solid lines show fits 
to the data, and dotted lines represent the results obtained by 
quasi-linear theory as given by Eq. (17). }
\end{figure}

\begin{figure}[h]
\figurenum{9}
\begin{center}
{\centerline{\hskip-0.15in\epsscale{1.0} \plotone{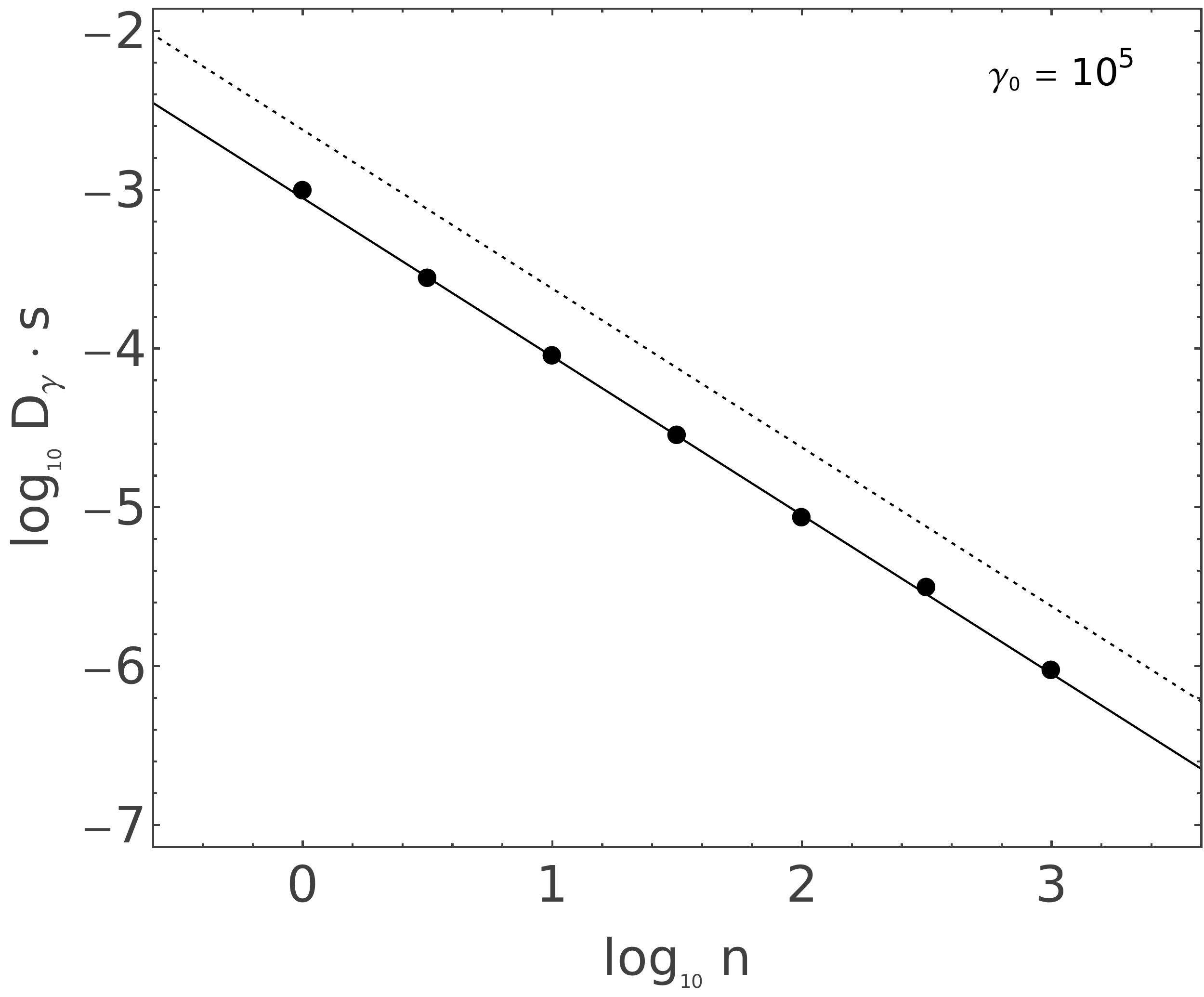} }}
\end{center}
\vskip -0.3in
\figcaption{Energy diffusion coefficients for $\gamma_0 = 10^5$
as a function of density $n$ for Experiments 1 and 9 --14 (isotropic turbulence). The uniform
magnetic field is $100\;\mu$G, $\eta = 1$, $\Gamma=5/3$, and 
$\lambda_{max}=0.1$ in every case. The solid line show the fit
to the data ($D_\gamma\propto n^{-1}$), and the dotted line represents the result obtained by 
quasi-linear theory as given by Eq. (17). }
\end{figure}

Figure 10 illustrates the energy dependence of
$D_\gamma$ for five values of background field strengths $B_0$. 
Since a greater magnetic field strength leads to a greater electric field 
strength, energy diffusion increases with field strength.  Again, the index $\alpha$ 
is not sensitive to $B_0$, but we find that
Equation (17) overestimates the energy diffusion for larger values of $\gamma_0$,
and underestimates the energy diffusion for smaller values of $\gamma_0$, although the transition
varies depending on the field strength.  
Figure 11 illustrates the relation between $D_\gamma$ and $B_0$ for the value
$\gamma_0 = 10^5$, where the fits from Figure 10 are used to obtain
the plotted data points.  
We note that the
scaling $D_\gamma \propto B_0^{2.50}$ obtained from our results differs from 
quasi-linear theory ($D_\gamma\propto B_0^{2.33}$).

Figure 12 illustrates the energy dependence of
$D_\gamma$ for three values of $\lambda_{max}$. 
Again, the index $\alpha$ 
is not sensitive to $\lambda_{max}$, and once again,
Equation (17) overestimates the energy diffusion for larger values of $\gamma_0$,
and underestimates the energy diffusion for smaller values of $\gamma_0$.
Figure 13 illustrates the relation between $D_\gamma$ and $\lambda_{max}$
for the value
$\gamma_0 = 10^5$, where the fits from Figure 12 are used to obtain
the plotted data points.  As was the case with the field, the scaling obtained ($D_\gamma \propto 
\lambda_{max}^{-0.47}$) differs from that predicted by
quasi-linear theory ($D_\gamma \propto 
\lambda_{max}^{-0.67}$).

Corresponding results for isotropic Kraichnan turbulence (derived
from experiments 5 and 21--26) are shown in Figures 14--17, 
and for isotropic Bohm turbulence (derived from experiments 7 and 
27--32) are shown in Figures 18--21. The analysis for these
cases mirrors that described above for Kolmogorov turbulence,
and the same general results are obtained.  

\begin{figure}[h]
\figurenum{10}
\begin{center}
{\centerline{\hskip-0.15in\epsscale{1.0} \plotone{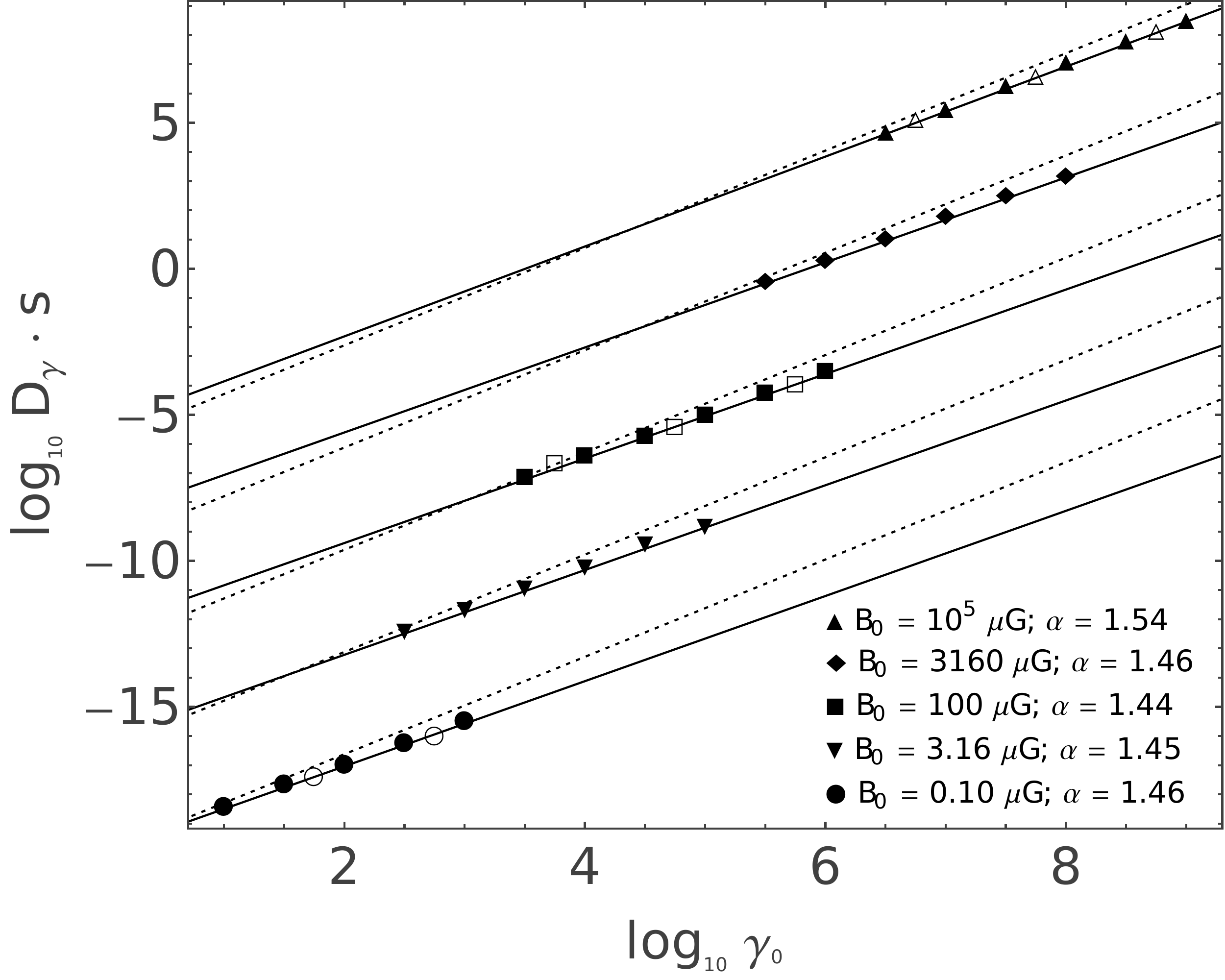} }}
\end{center}
\vskip -0.3in
\figcaption{Energy diffusion coefficients as a function of particle
Lorentz factor for Experiments 1 and 15--18, which sample various 
background field strengths $B_0$. The particle density $n = 100$ cm$^{-3}$,
$\eta = 1$, $\Gamma=5/3$, and $\lambda_{max}=0.1$ in every case. 
Solid shapes denote values for isotropic turbulence,
and open shapes denote values for anisotropic turbulence. 
Solid lines show fits 
to the data, and dotted lines represent the results obtained by 
quasi-linear theory as given by Eq. (17).  }
\end{figure}

\begin{figure}
\figurenum{11}
\begin{center}
{\centerline{\hskip-0.15in\epsscale{1.0} \plotone{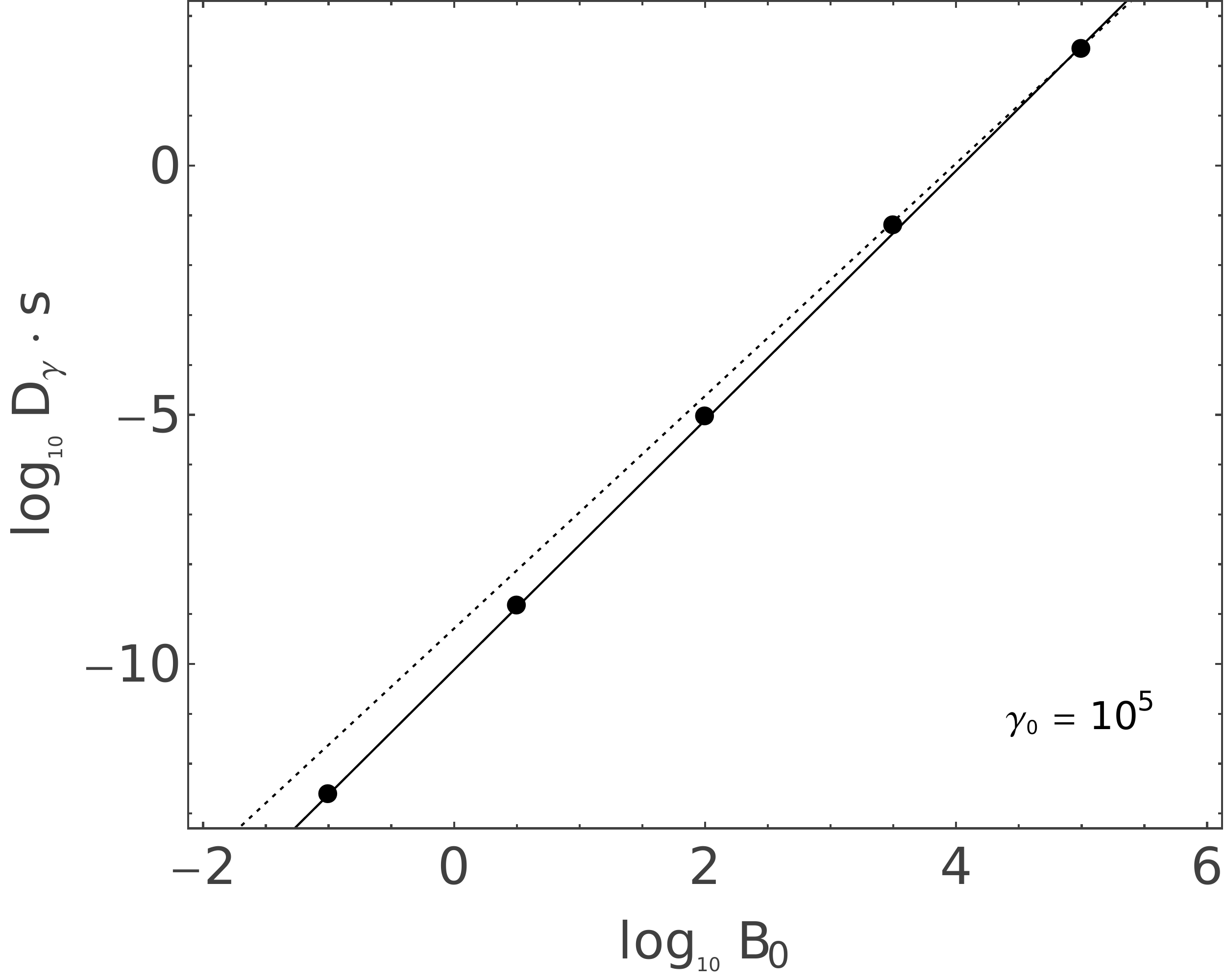} }}
\end{center}
\vskip -0.3in
\figcaption{Energy diffusion coefficients for $\gamma_0 = 10^5$
as a function of background field strength $B_0$
for Experiments 1 and 15--18 (isotropic turbulence). The particle density $n = 100$ cm$^{-3}$,
$\eta = 1$, $\Gamma=5/3$, and $\lambda_{max}=0.1$ in every case. 
The solid line shows the fit 
to the data ($D_\gamma\propto B_0^{2.50}$), and the dotted line 
represents the result obtained by 
quasi-linear theory as given by Eq. (17).}
\end{figure}

\begin{figure}
\figurenum{12}
\begin{center}
{\centerline{\hskip-0.15in\epsscale{1.0} \plotone{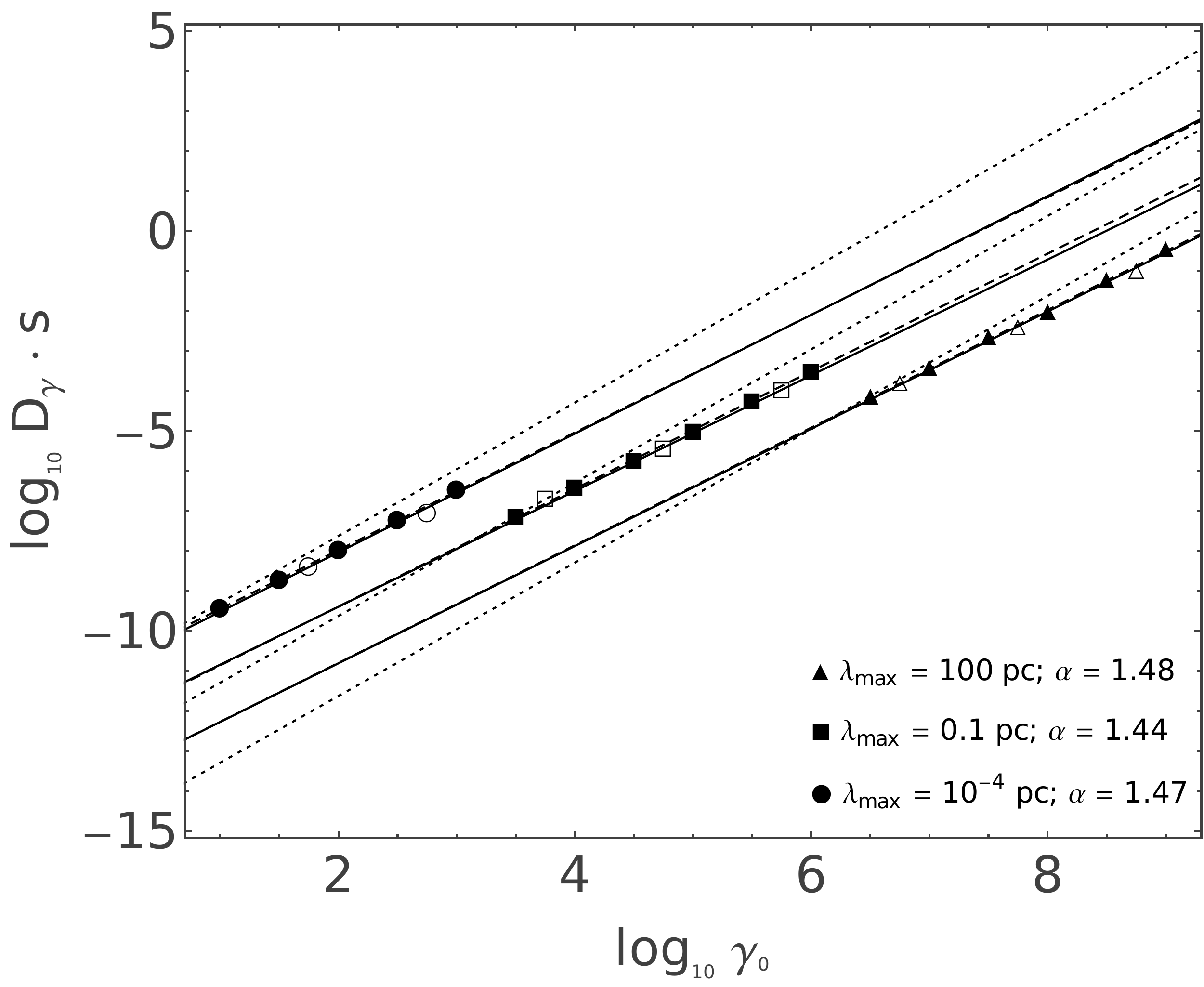} }}
\end{center}
\vskip -0.3in
\figcaption{Energy diffusion coefficients as a function of particle
Lorentz factor for Experiments 1 and 19--20, which sample various 
values of  $\lambda_{max}$. The particle density $n = 100$ cm$^{-3}$,
background field strength $B_0 = 100\mu$G, 
$\eta = 1$, and $\Gamma=5/3$ in every case. 
Solid shapes denote values for isotropic turbulence,
and open shapes denote values for anisotropic turbulence. 
Solid lines show fits 
to the data, and dotted lines represent the results obtained by 
quasi-linear theory as given by Eq. (17). }
\end{figure}

\begin{figure}
\figurenum{13}
\begin{center}
{\centerline{\hskip-0.15in\epsscale{1.0} \plotone{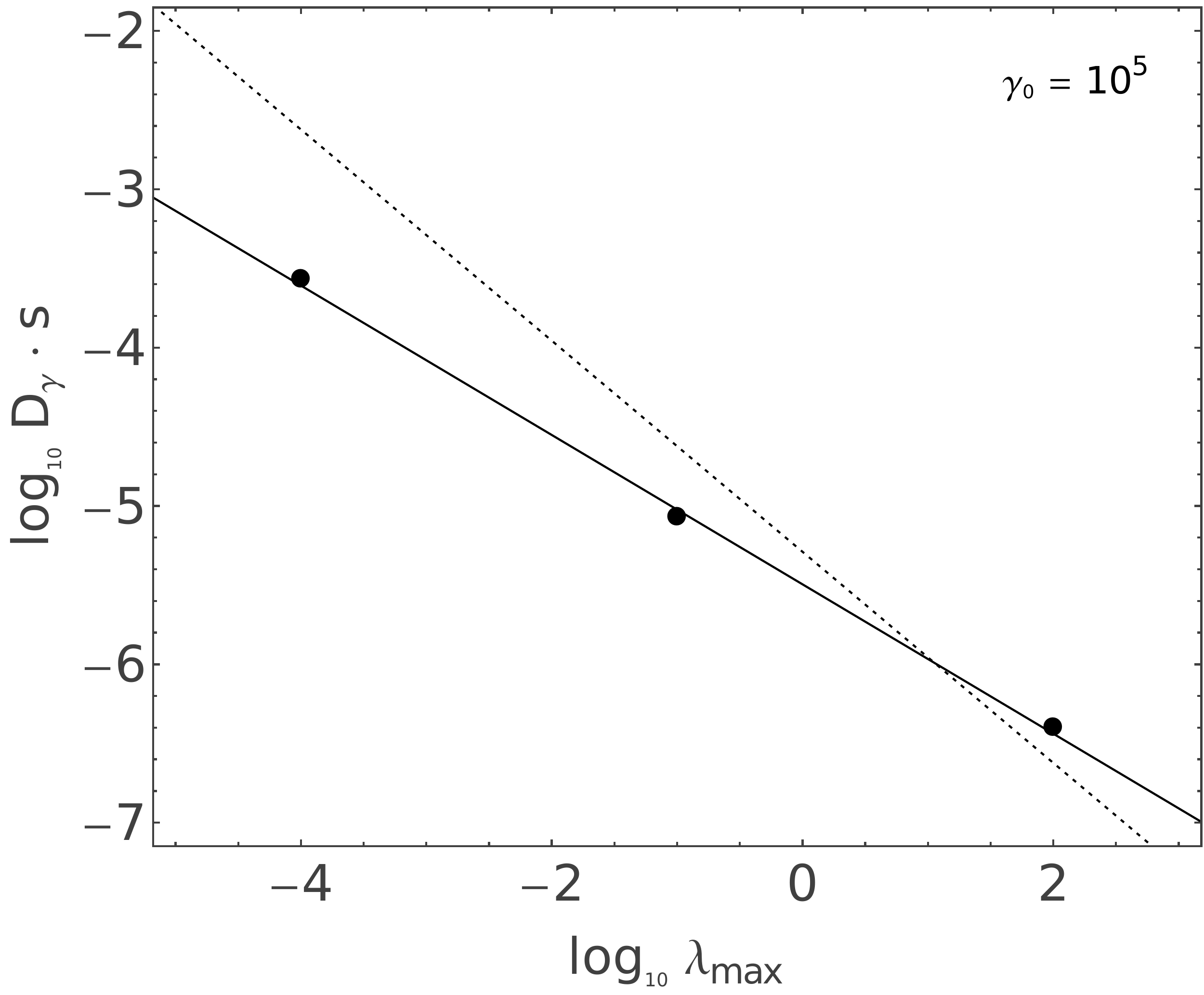} }}
\end{center}
\vskip -0.3in
\figcaption{Energy diffusion coefficients for $\gamma_0 = 10^5$
as a function of $\lambda_{max}$
for Experiments 1 and 19--20 (isotropic turbulence). The particle density $n = 100$ cm$^{-3}$,
background field strength $B_0 = 100\mu$G,
$\eta = 1$, and $\Gamma=5/3$ in every case. 
The solid line shows the fit 
to the data ($D_\gamma\propto \lambda_{max}^{-0.47}$), and the dotted line represents the result obtained by 
quasi-linear theory as given by Eq. (17).}
\end{figure}
\begin{figure}

\figurenum{14}
\begin{center}
{\centerline{\hskip-0.15in\epsscale{1.0} \plotone{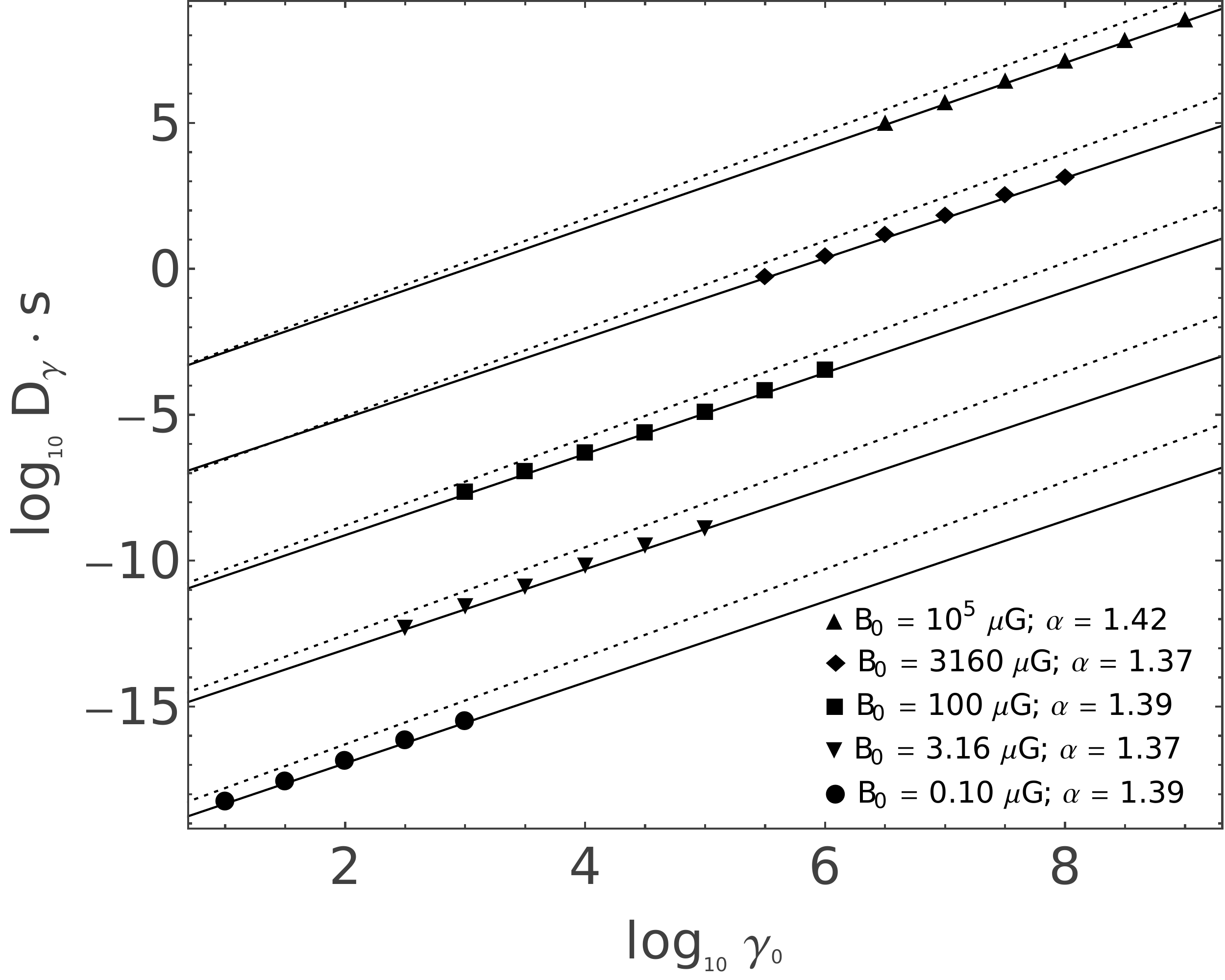} }}
\end{center}
\vskip -0.3in
\figcaption{Same as Figure 10, but for $\Gamma = 3/2$ and Experiments
5 and 21--24.
}
\end{figure}

\begin{figure}
\figurenum{15}
\begin{center}
{\centerline{\hskip-0.15in\epsscale{1.0} \plotone{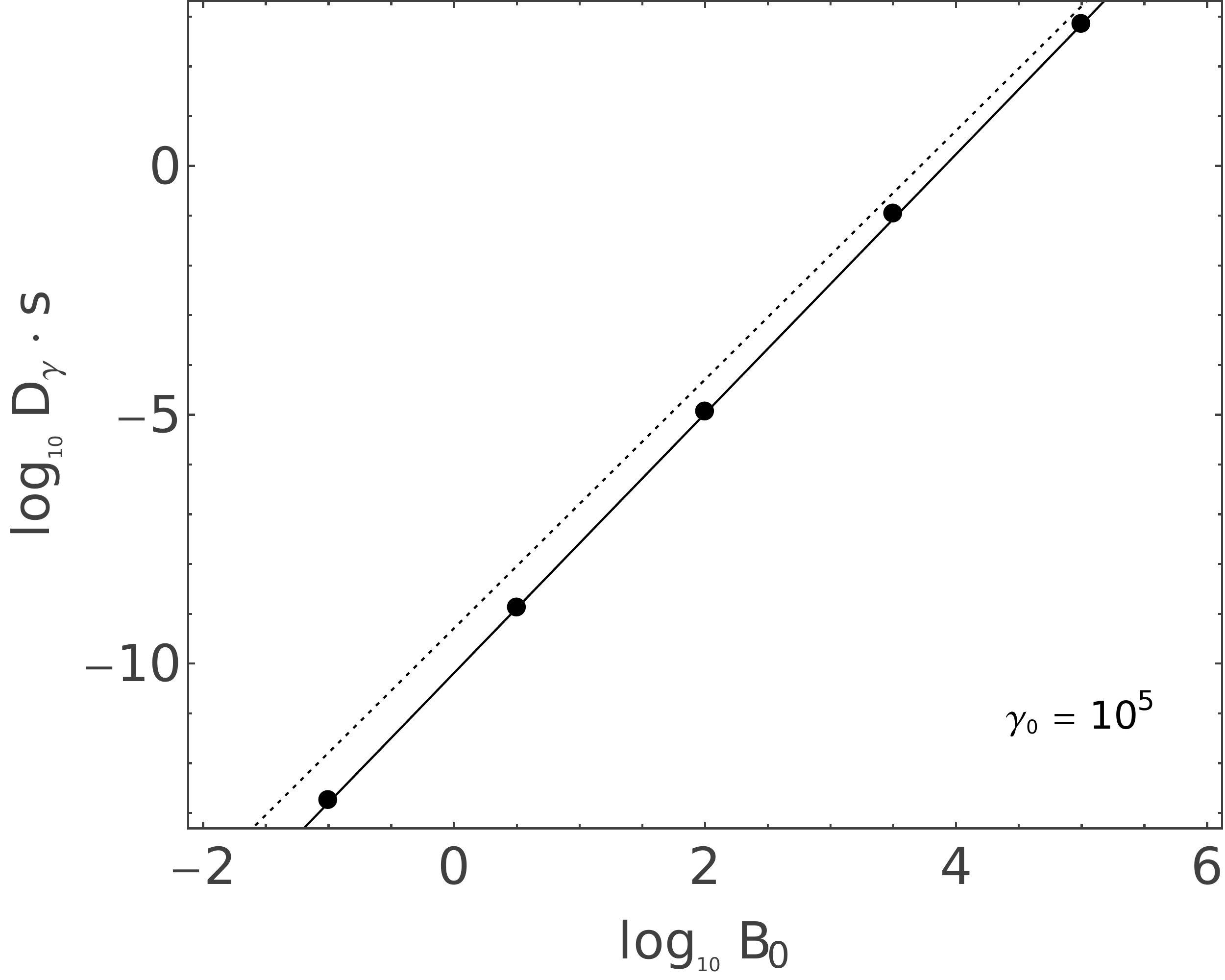} }}
\end{center}
\vskip -0.3in
\figcaption{Same as Figure 11, but for $\Gamma = 3/2$ and Experiments
5 and 21--24.  The fit to the data (as shown by the solid line) yields
$D_\gamma\propto B_0^{2.61}$.}
\end{figure}

\begin{figure}
\figurenum{16}
\begin{center}
{\centerline{\hskip-0.15in\epsscale{1.0} \plotone{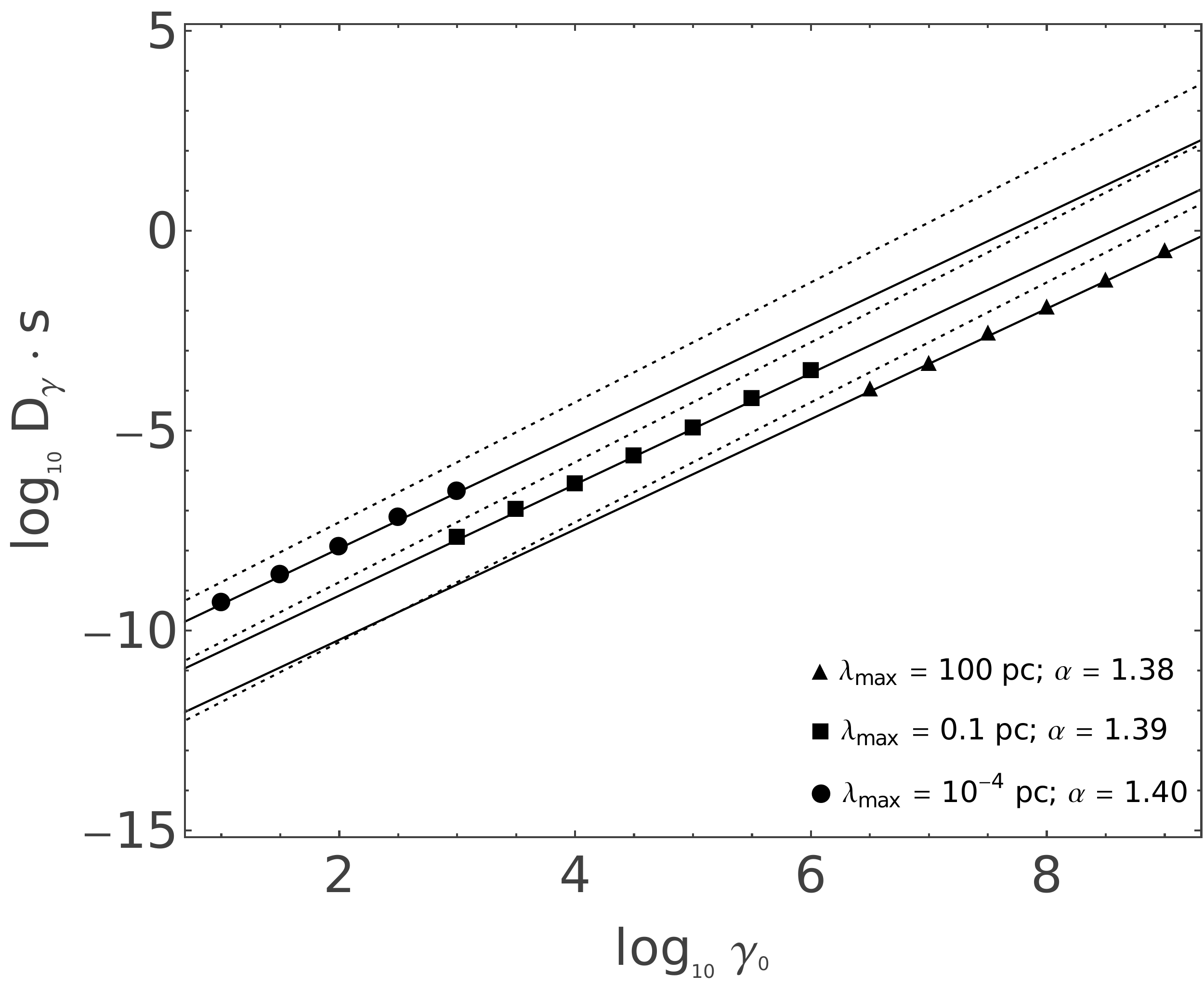} }}
\end{center}
\vskip -0.3in
\figcaption{Same as Figure 12, but for $\Gamma = 3/2$ and Experiments
5 and 25--26.}
\end{figure}

\begin{figure}
\figurenum{17}
\begin{center}
{\centerline{\hskip-0.15in\epsscale{1.0} \plotone{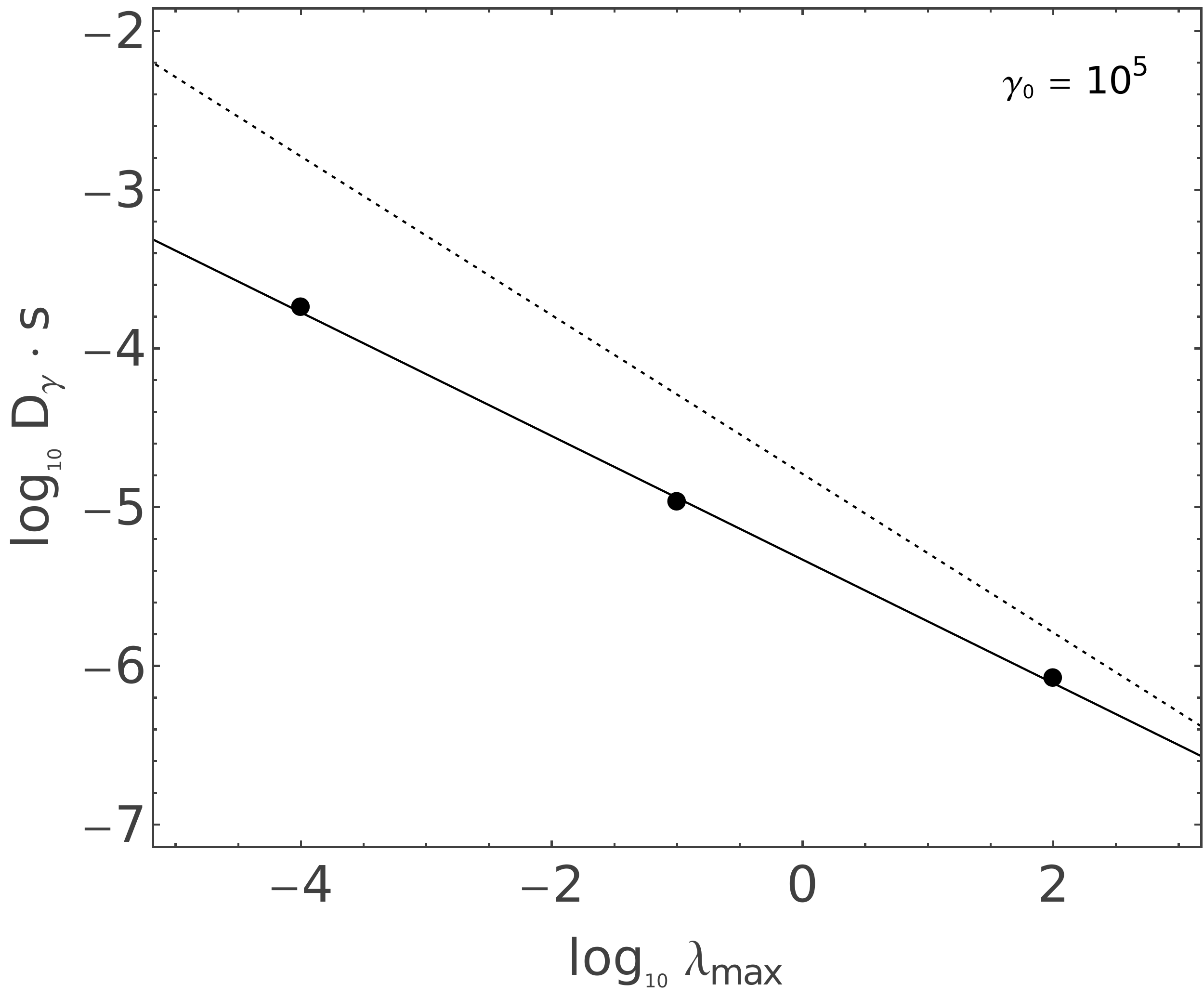} }}
\end{center}
\vskip -0.3in
\figcaption{Same as Figure 13, but for $\Gamma = 3/2$ and Experiments
5 and 25--26.  The fit to the data (as shown by the solid line) yields
$D_\gamma\propto \lambda_{max}^{-0.39}$.}
\end{figure}

\begin{figure}
\figurenum{18}
\begin{center}
{\centerline{\hskip-0.15in\epsscale{1.0} \plotone{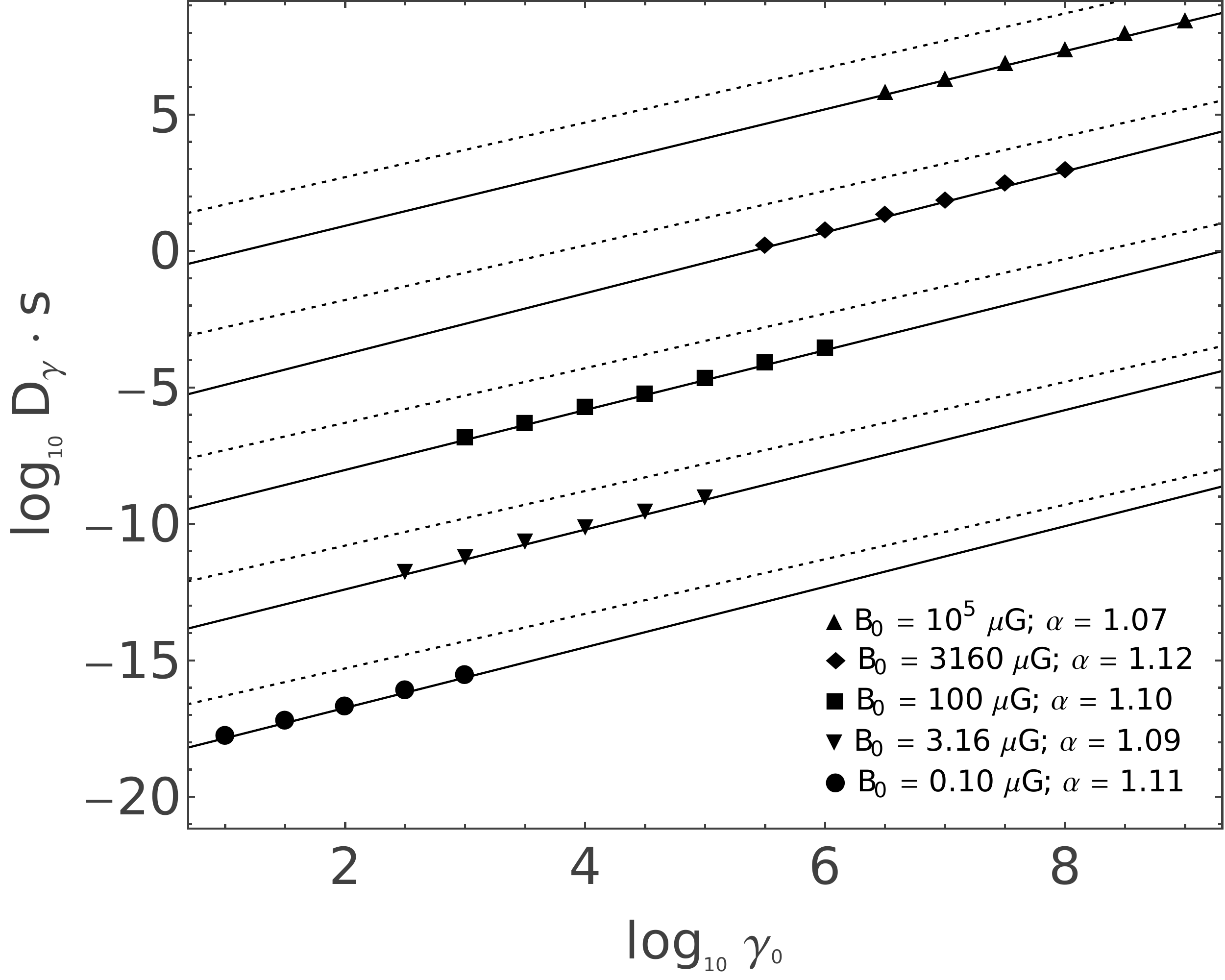} }}
\end{center}
\vskip -0.3in
\figcaption{Same as Figure 10, but for $\Gamma = 1$ and Experiments
7 and 27--30.}
\end{figure}

\begin{figure}
\figurenum{19}
\begin{center}
{\centerline{\hskip-0.15in\epsscale{1.0} \plotone{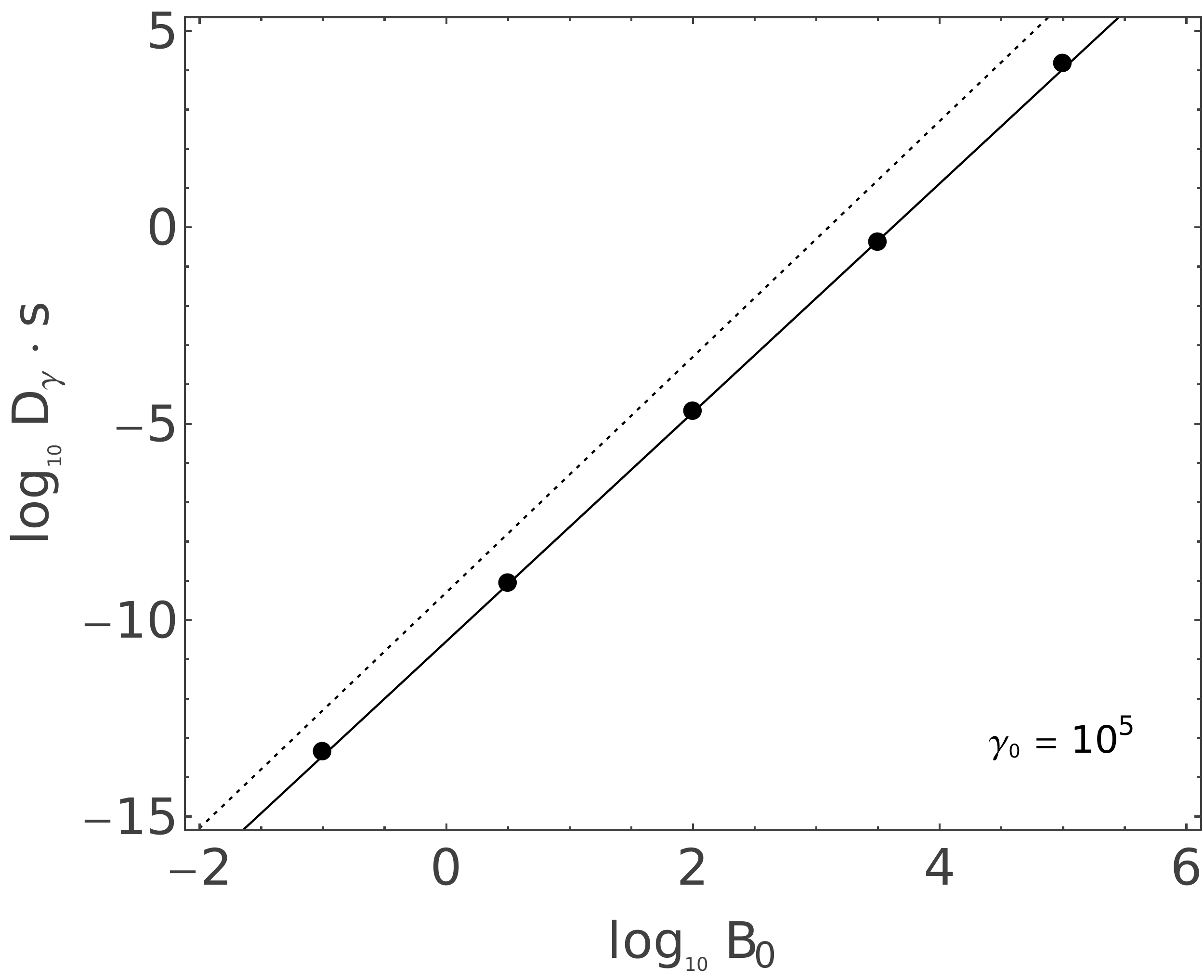} }}
\end{center}
\vskip -0.3in
\figcaption{Same as Figure 11, but for $\Gamma = 1$ and Experiments
7 and 27--30. The fit to the data (as shown by the solid line) yields
$D_\gamma\propto B_0^{2.91}$.}
\end{figure}

\begin{figure}
\figurenum{20}
\begin{center}
{\centerline{\hskip-0.15in\epsscale{1.0} \plotone{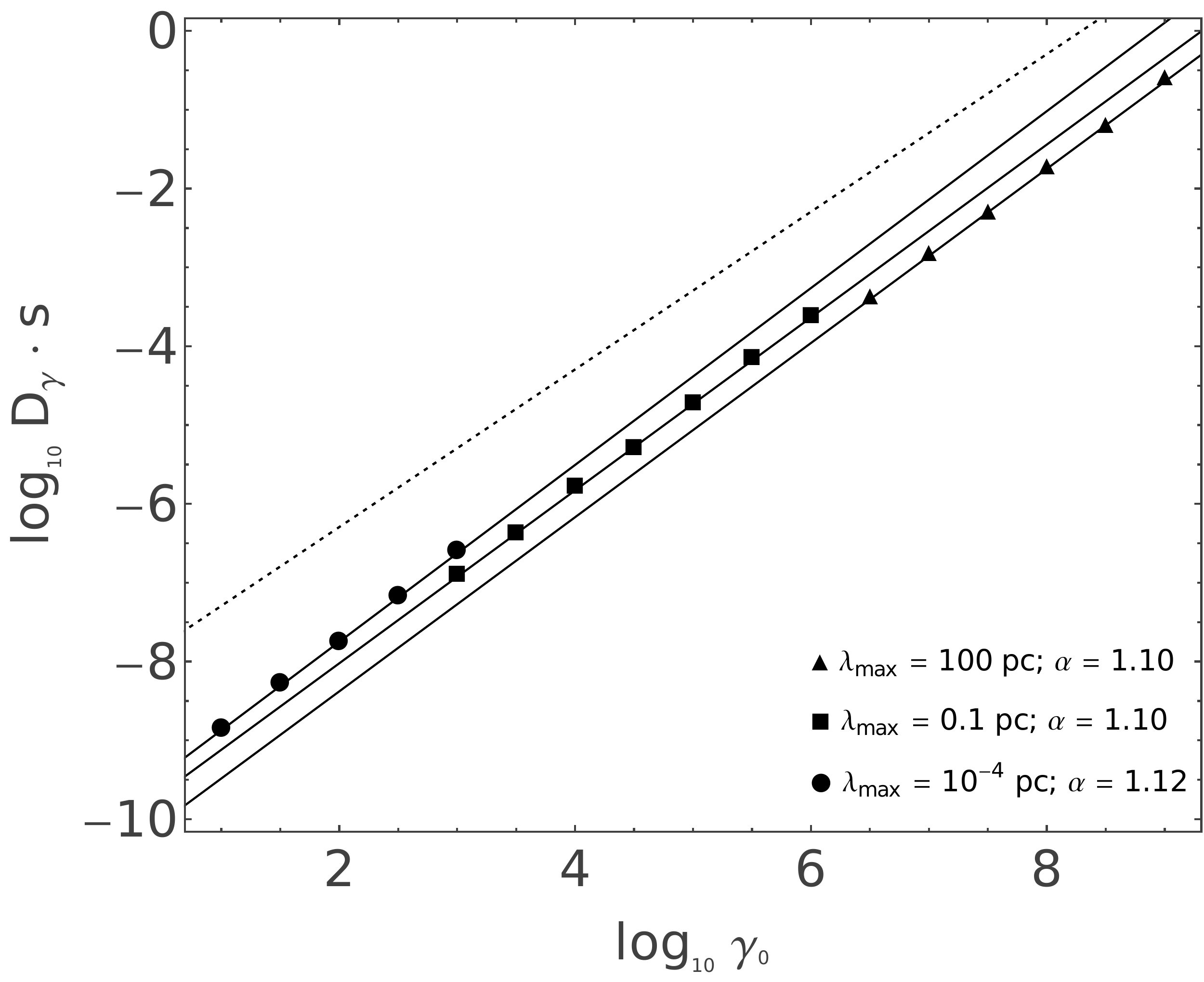} }}
\end{center}
\vskip -0.3in
\figcaption{Same as Figure 12, but for $\Gamma = 1$ and Experiments
7 and 31--32.}
\end{figure}

\begin{figure}[h]
\figurenum{21}
\begin{center}
{\centerline{\hskip-0.15in\epsscale{1.0} \plotone{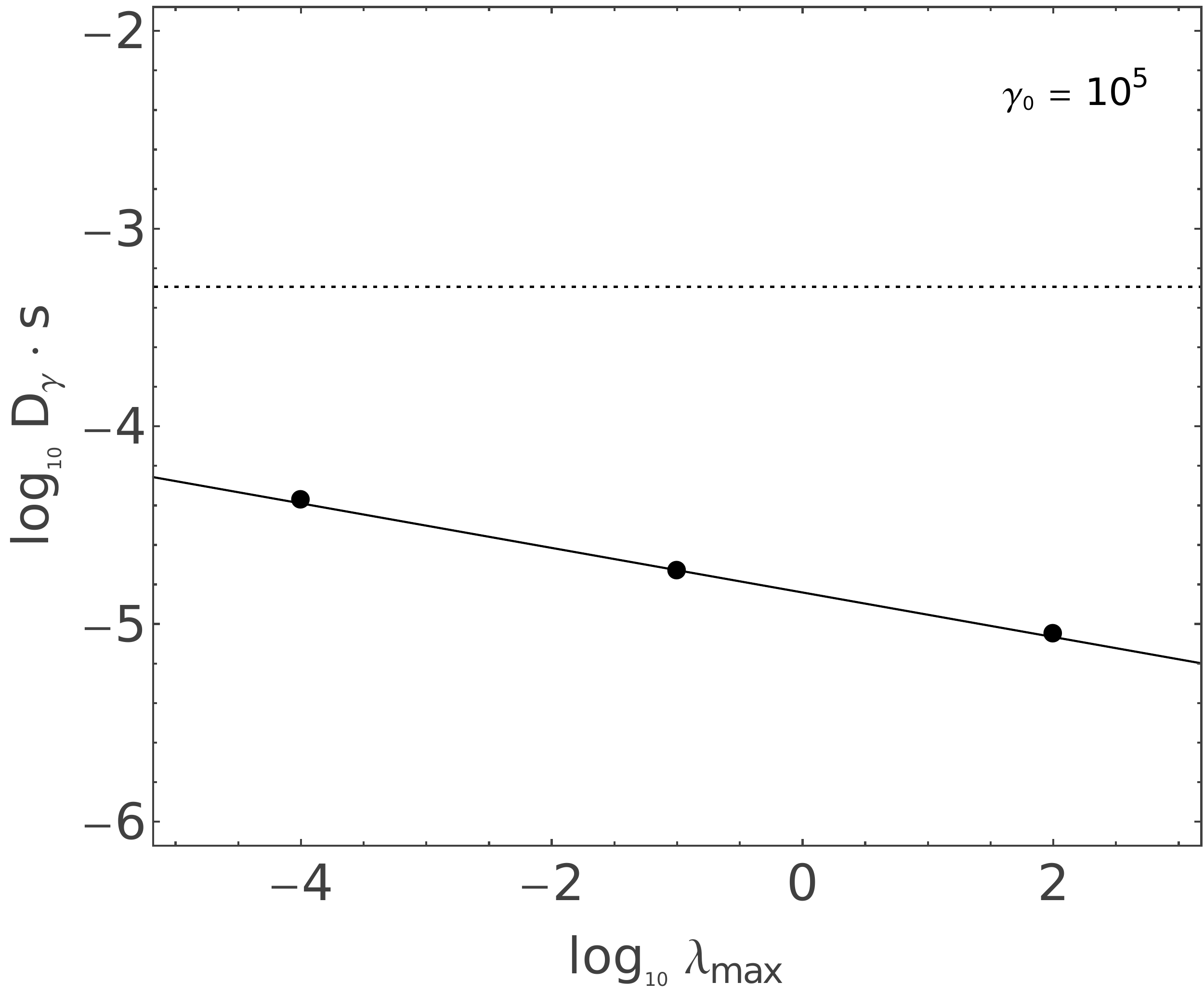} }}
\end{center}
\vskip -0.3in
\figcaption{Same as Figure 13, but for $\Gamma = 1$ and Experiments
7 and 31--32. The fit to the data (as shown by the solid line) yields
$D_\gamma\propto \lambda_{max}^{-0.11}$.}
\end{figure}

Our results indicate that simple scaling laws of the form
$D_\gamma = D_{\gamma 0} \,\gamma^\alpha$, where
\be
D_{\gamma 0}= D_0 \left({n\over 1\,{\rm cm}^{-3}}\right)^{-1}
 \left({B_0\over 1\,\mu{\rm G}}\right)^\delta
 \left({\lambda_{max}\over 1 \,{\rm pc}}\right)^\kappa ,
\ee
can be used to obtain
values of the energy diffusion coefficient over a wide range of parameters
that pertain to turbulent interstellar and intergalactic environments, so long
as the particle gyration radius $\lambda_{min} \ll R_g \ll \lambda_{max}$.
In addition, the same expressions can be used to describe the isotropic and 
anisotropic cases in the limit of strong turbulence ($\eta\approx 1$).

We obtain values of $D_0$,  $\alpha$, $\delta$ and $\kappa$ in the strong turbulent limit 
($\eta = 1$) for the three turbulence profiles 
considered in our work by using the results presented above.
Specifically, final values of $D_0$ and $\alpha$, along with their
$1 \sigma$ errors, are obtained by finding the mean and standard deviations
of the $D_0$ and $\alpha$ values from the corresponding experiments listed in 
Table 1.   The values of $\delta$ and $\kappa$ are obtained from the fits to the
data shown in figure~11 and figure~13 for $\Gamma = 5/3$,  figure~15 and figure~17 
for $\Gamma = 3/2$, and  figure~19 and figure~21 for $\Gamma = 1$. 
The results are summarized in Table 2.

\begin{deluxetable}{cccccc}
\tablecolumns{6}
\tablewidth{0pc}
\tablecaption{Fitting parameters}
\tablehead{
\colhead{$\Gamma$}  &&
\colhead{$D_0 \, ({\rm s}^{-1})$}   & \colhead{$\alpha$}& \colhead{$\delta$}&\colhead{$\kappa$} }
\startdata
5/3   && $(1.6\pm 0.8)\times 10^{-16}$ & $1.47\pm0.02$  & 2.50 & -0.47\\
3/2   && $(3.7\pm 1.7)\times 10^{-16}$ & $1.39\pm0.02$  & 2.61 & -0.39\\
1   && $(7.6\pm 2.8)\times 10^{-15}$ & $1.10\pm0.02$  & 2.91 & -0.11\\
\enddata
\end{deluxetable}

We conclude our analysis by applying the results of this work toward obtaining
estimates of the acceleration time $\tau_{acc} \equiv \gamma^2 / D_\gamma$
required to energize protons up to energies of 1 TeV in molecular cloud environments.
To keep this analysis as simple as possible, we assume 
that the magnetic field scales with density as given by Eq. (1).
We also assume that the maximum turbulence wavelength $\lambda_{max}$ 
scales as the cloud size.   Following Larson's law (Larson 1981), we obtain
the relation
\be
\lambda_{max}  = 1600 \,{\rm pc} \left({n_{H_2}\over 1 \,{\rm cm}^{-3}}\right)^{-0.91}\,.
\ee 
A plot of the acceleration time $\tau_{acc}$ as a function of molecular hydrogen density $n_{H_2}$
is shown in Figure 22 for Kolmogorov, Kraichnan, and Bohm turbulence.  Given that clouds
are not expected to last for more than $\sim 10$ Myrs, TeV cosmic ray production from 
stochastic acceleration of turbulent magnetic fields can clearly be ruled our for
normal molecular cloud environments.   We note, however, that the molecular
clouds near the galactic center have fairly extreme environments.  As shown in 
Fatuzzo \& Melia (2012), the acceleration 
times is the GC environment are considerably shorter, as illustrated by 
the solid circle (inter cloud region at the GC) and solid square (molecular
cloud at the GC) shown in Figure 22. 

\begin{figure}
\figurenum{22}
\begin{center}
{\centerline{\hskip-0.15in\epsscale{1.0} \plotone{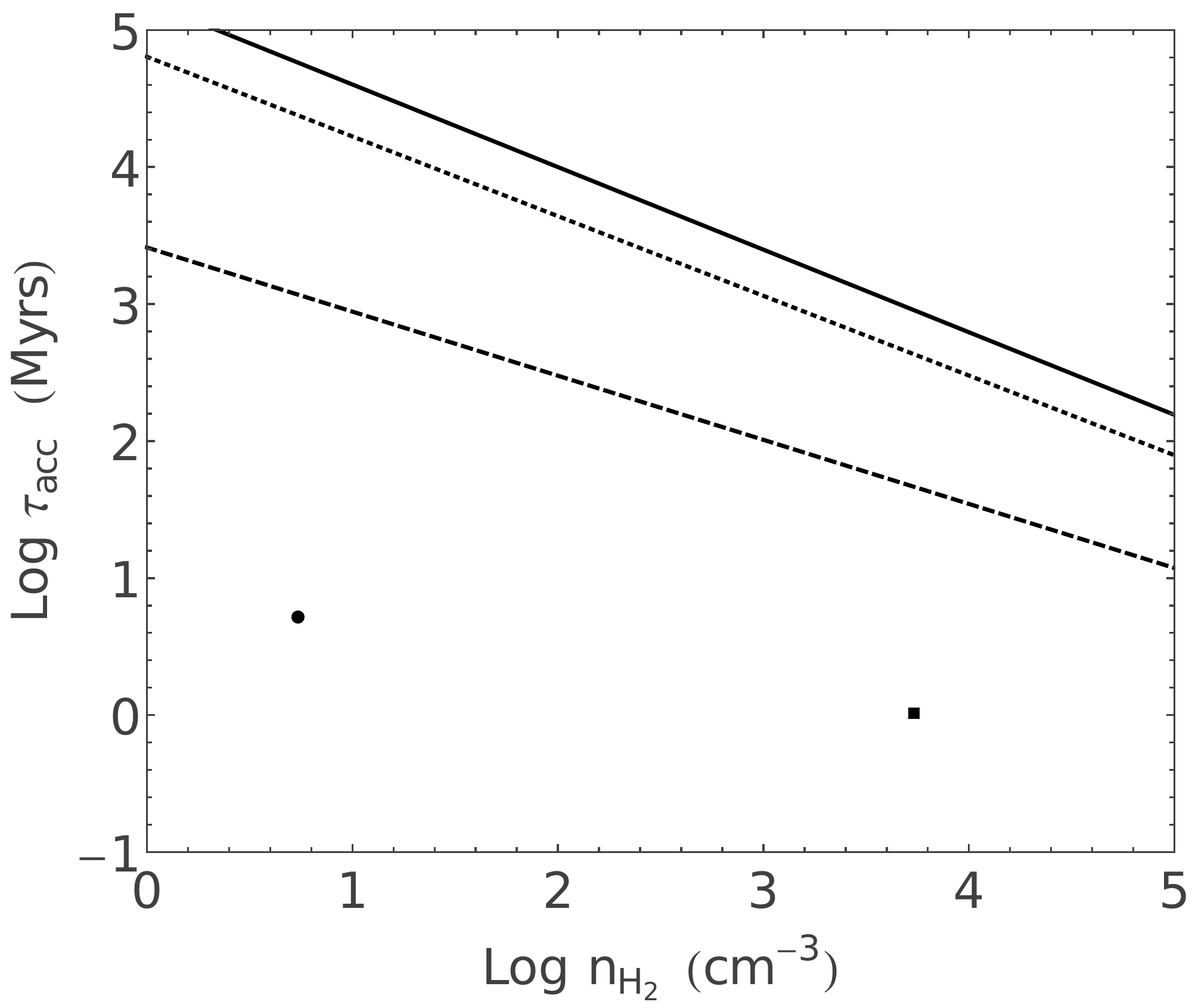} }}
\end{center}
\vskip -0.3in
\figcaption{The acceleration time $\tau_{acc} \equiv \gamma^2 / D_\gamma$ as a function
of molecular hydrogen density for molecular cloud environments.  The magnetic field is
assumed to scale with density as per Eq. [1], and $\lambda_{max}$ is set equal to the
size of the molecular cloud, where Larson's Law is then used to relate 
$\lambda_{max}$ to the density $n_{H_2}$.  The solid curve denotes the results obtained 
for $\Gamma = 5/3$, the dotted curve denotes the results obtained for $\Gamma = 3/2$,
and the dashed curve denotes the results obtained for $\Gamma = 1$.
For comparison, the solid circle and solid square represent the acceleration times in the 
inter cloud region and molecular clouds at the GC, respectively (Fatuzzo \& Melia 2012).
}
\end{figure}

\section{Conclusions}
We have used a detailed numerical simulation to determine the spatial and spectral
profiles of cosmic rays diffusing with arbitrary energy through turbulent media
characterized by a broad range of magnetic fields, turbulence strength, fluctuation
size and ambient particle density, for both isotropic and anisotropic turbulence. 
This study has expanded considerably from the
initial attempt at simulating the behavior of relativistic particle propagation through the
galactic-center environment, where they encounter a variety of physical conditions,
inside and outside of the molecular gas in that region. Our goal throughout this exercise has
been to avoid using ``standard" techniques, e.g., quasi-linear theory or the diffusion
equation, all of which are often subject to unknown factors that delimit the applicability
of these approaches to real systems (but see also Nayakshin \& Melia 1998; Wolfe \& Melia
2006). Indeed, one of the prinicpal benefits of the technique we have developed in this 
work is an accurate determination of the spatial and energy diffusion coefficients that 
in turn may be used in these other approaches without the need to guess or estimate their 
normalization and energy dependence.

As the sensitivity and spectral range of high-energy observatories continue to
improve, the need to accurately simulate the propagation of relativistic particles
through turbulent media arises in an ever increasing range of environments, from
the interstellar medium, to compact accretion regions surrounding supermassive
black holes (such as Sgr A* at the galactic center), to the hot intracluster gas, and
in even more exotic structures, such as the Fermi bubbles straddling the center of
the Milky Way (Crocker \& Aharonian 2011). However, the physical conditions
characterizing these regions change considerably from one environment to the
next. For example, the magnetic intensity may be as large as several Gauss
near accreting black holes, but smaller than $0.01\;\mu$G in the intergalactic
medium. We now know that quasi-linear theory is only approximately valid
even in weakly turbulent environments, let alone regions where the turbulent
magnetic energy is comparable to, or bigger than, the underlying uniform component.
Worse, it is often difficult to estimate the absolute value of the diffusion coefficients
without resorting to observational data which, however, are sometimes difficult to get
(as in the case of the giant radio lobes in FR II galaxies).

These are among the reasons we have embarked on this type of investigation,
to develop methods of handling the great diversity of physical conditions
encountered by cosmic rays propagating through high-energy emitting
environments. Previously (Fatuzzo \& Melia 2010, 2012b), we reported
some results of this study pertaining to the spatial diffusion coefficients.
We found that the spatial diffusion of particles through turbulent fields is not
sensitive to the minimum wavelength of the fluctuations, so long as the
particle's radius of gyration exceeds $\lambda_{min}$. For a given
environment, as defined by $B_0$ and $v_A$, the diffusion process is
thus dependent upon the maximum turbulence wavelength $\lambda_{max}$,
the turbulent field strength, as characterized by $\eta$, and the turbulence
spectrum, as characterized by the spectral index $\Gamma$.

We also noted that quasi-linear theory does not appear to be valid
in the strong turbulence limit (see also O'Sullivan et al. 2009). We
therefore investigated how the energy diffusion coefficient depends
upon $\lambda_{max}$ and $\eta$ for Kolmogorov ($\Gamma = 5/3$) turbulence,
and found that the energy diffusion coefficients could be characterized
as $D_\gamma \propto\lambda_{max}^{-0.47}$. This behavior
is not consistent with quasi-linear theory, which instead predicts that
$D_\gamma \propto \lambda_{max}^{-0.67}$ for Kolmogorov
turbulence in the strong turbulence ($\eta \ga 1$) limit. However,
we also found that $D_\gamma \propto \eta^{1.2}$ in both the weak
and strong turbulence limits.

But clearly, this initial sampling of the complex behavior of $D_\gamma$
under a variety of physical conditions is far from satisfactory. The purpose
of the present paper has been to complete this work, finding scaling
relations that one may use to calculate $D_\gamma$ under most
conditions of interest, for all practical ranges of magnetic intensity,
turbulence strength, ambient particle density, fluctuation size, and
turbulence spectrum.

The empirical relations useful for this purpose have all been presented in
\S5. Broadly speaking, we have found that insofar as the energy diffusion
coefficient $D_\gamma$ is concerned, quasi-linear theory predicts its correct
energy dependence only for very weak, isotropic turbulence (i.e., $\eta\la 0.01$).
These predictions deviate substantially for $\eta\sim 1$, particularly for
turbulence spectral indeces $\Gamma >1$, such as Kolmogorov turbulence,
which seems to be prevalent in many diverse environments. For example,
for the physical conditions one encounters at the galactic center (i.e.,
$B\sim 1-10^3\;\mu$G and $n\sim 1-10^3$ cm$^{-3}$), the actual index
characterizing the dependence of $D_\gamma$ on $\gamma$ may be
as small as $\sim 1.4$ instead of the predicted value $\sim 1.7$.

In addition, our results indicate that there is no difference in the energy
diffusion of particles between isotropic turbulence and the anisotropic turbulence
profiles predicted by Goldreich \& Sridhar (1995) so long as the turbulence is
strong ($\delta B \sim B$).  On the other hand, our results indicate that anisotropic
weak turbulence is considerably less effective in energetically scattering particles,
consistent with the results of Chanrdran (2000).

Needless to say, if one is attempting to interpret the GeV Fermi spectrum
of, say, the Fermi bubbles, in terms of an underlying population of
cosmic rays, deviations from quasi-linear theory are rather
critical, since the inferred particle distribution differs considerably
from its injection point to where it emits the radiation, and the difference
will be interpreted incorrectly with the inaccurate energy dependence
predicted by these other techniques.

And not being sure of the normalization of $D_\gamma$ has its own
challenges. For one thing, it is not possible to say anything definitive
about the overall power being generated by the acceleration of
these cosmic rays, which speaks directly to the mechanism associated
with the relativistic particle injection, or even to the required density
of dark-matter particles, if these cosmic rays are produced via
dark-matter decays and collisions. But with our approach, it is not
necessary to estimate the normalization of $D_\gamma$, because
its absolute value is determined self-consistently from the statistical
aggregate of numerous individual particle trajectories.

We have found that quasi-linear theory provides an acceptable
estimate of the normalization of $D_\gamma$ at $\sim 1$ TeV
energies, but can deviate considerably at lower energies,
especially in the GeV range, and at energies exceeding $10-100$
TeV. A large factor responsible for these differences is the incorrect
energy dependence predicted for $D_\gamma$. Obviously, if
the normalization is adequate at $\sim 1$ TeV, the incorrect energy
index will cause deviations at lower and higher energies.

Most importantly, however, our analysis has provided a method of
determining not only the dependence of $D_\gamma$ on $\gamma$,
but also its absolute value without the need to normalize it from the
data. Having said this, one is not completely free of ambiguity,
since one must still have an accurate estimate of the physical
conditions, i.e., the magnetic field, the ambient density, and other
characteristics that determine the state of the medium though which
the cosmic rays are propagating. Fortunately, these conditions are
easier to measure than the diffusion coefficients themselves, and
in a more sophisticated use of our technique, in which MHD turbulence
is simulated numerically rather than via the simple Kolmogorov or
Bohm scaling relations, one can approach a level of realism not
available to any of the other methods.

We look forward to the application of the scaling relations we
have presented in this paper across a broad range of physical
environments, allowing us to study high-energy sources with
a level of accuracy commensurate with the detailed measurements
now being made by the ever improving suite of space-based and
ground-based observatories.

\acknowledgments

This work was supported by Xavier University through the Hauck Foundation, 
and by ONR grant N00014-09-C-0032 at the University of Arizona.  







\begin{thebibliography}

\bibitem[]{A11} Adriani, O. et al. 2011, Science, Volume 332, 6025, 69
\bibitem[]{Aha06} Aharonian, F. et al. 2006, Nature, 439, 695
\bibitem[]{Bal07} Ballantyne, D., Melia, F., \& Liu, S. 2007, ApJ Letters, 657, L13
\bibitem[]{C06} Casse, F., Lemoine, M., \& Pelletier, G. 2002, Phys. Rev. D, 65, 023002
\bibitem[]{Ch00} Chandran, B. D. G. 2000, Phys. Rev. Lett, 85, 4656
\bibitem[]{C00} Cho, J. \& Vishniac, E. T. 2000, ApJ, 539, 273
\bibitem[]{C97} Coker, R. F. \& Melia, F. 1997, ApJ Lett, 488, L149
\bibitem[]{C11} Crocker, R. M. \& Aharonian, F. 2011, Phys. Rev. Lett.,106, 101102
\bibitem[]{C99} Crutcher, R. M. 1999, ApJ, 520, 706
\bibitem[]{F97} Falcke, H. \& Melia, F. 1997, ApJ, 479, 740
\bibitem[]{F06} Fatuzzo, M., Adams, F. C. \& Melia, F. 2006, ApJ Letters, 653, L49
\bibitem[]{F10} Fatuzzo, M., Melia, F., Wommer, E. \& Adams, F. 2010, ApJ, 725, 515
\bibitem[]{F11} Fatuzzo, M. \& Melia, F. 2011, MNRAS, 410, L23
\bibitem[]{F12a} Fatuzzo, M. \& Melia, F. 2012a, ApJ Letters, 757, L16
\bibitem[]{F12b} Fatuzzo, M. \& Melia, F. 2012b, ApJ, 750, 21
\bibitem[]{F09} Fraschetti, F. \& Melia, F. 2008, MNRAS, 391, 1100
\bibitem[]{G84} Giacalone, J. \& Jokipii, J. R. 1994, ApJ Letters, 430, L137
\bibitem[]{GS95} Goldreich, P. \& Sridhar, S., ApJ, 438, 763
\bibitem[]{K99} Kowalenko, V. \& Melia, F. 1999, MNRAS, 310, 1053
\bibitem[]{K94} Kronberg, P. P. 1994, Reports on Progress in Physics, 57, 325
\bibitem[]{L91} Lada, E. A., Bally, J., \& Stark, A. A.1991, ApJ, 368, 432
\bibitem[]{L81} Larson, B. L. 1981, MNRAS, 194, 809
\bibitem[]{L89} Lis, D. C. \& Goldsmith, P. F. 1989, ApJ, 337, 704
\bibitem[]{L90} Lis, D. C. \& Goldsmith, P. F. 1990, ApJ, 356, 195
\bibitem[]{L01} Liu, S. \& Melia, F. 2001, ApJ Letters, 561, L77
\bibitem[]{L06a} Liu, S., Melia, F., Petrosian, V. \& Fatuzzo, M. 2006a, ApJ Letters, 647, L1099
\bibitem[]{L06b} Liu, S., Petrosian, V., Melia, F. \& Fryer, C. L. 2006b, ApJ, 648, 1020
\bibitem[]{L12} Lynn, J. W., Parrish, I. J., Quataert, E., \& Chandran,  B. D. G. 2012, ApJ, 758, 78
\bibitem[]{L13} Lynn, J. W.,  Quataert, E., Chandran,  B. D. G., \& Parrish, I. J. 2013, ApJ, 777, 128
\bibitem[]{M97} Markoff, S., Melia, F. \& Sarcevic, I. 1997, ApJ Lett, 489, 47L
\bibitem[]{M99} Markoff, S., Melia, F. \& Sarcevic, I. 1999, ApJ, 522, 870
\bibitem[]{M07} Melia, F. 2007, The Galactic Supermassive Black Hole, Princeton University Press (NY)
\bibitem[]{MC99} Melia, F. \& Coker, R. F. 1999, ApJ, 511, 750
\bibitem[]{M93} Misra, R. \& Melia, F. 1993, ApJ Lett, 419, L25
\bibitem[]{N98} Nayakshin, S. \& Melia, F. 1998, ApJS, 114, 269
\bibitem[]{R94} Ruffert, M. \& Melia, F. 1994, A\&A Lett, 288, L29
\bibitem[]{Oxx} O'Sullivan, S., Reville, B., \& Taylor, A. M. 2009, MNRAS, 400, 248
\bibitem[]{P98} Paglione, D. T. A., Jackson, J. M., Bolatto, A. D., \& Heyer, M. H. 1998, ApJ, 493, 680
\bibitem[]{S89} Schlickeiser, R. 1989, ApJ, 336, 243
\bibitem[]{W88} Wefel, J. 1988, in Genesis and propagation of cosmic rays; Proceedings of the NATO
Advanced Study Institute (A88-37036 14-93), 1 
\bibitem[]{W06} Wolfe, B. \& Melia, F. 2006, 638, 125
\bibitem[]{wom08} Wommer, E., Melia, F., \& Fatuzzo, M. 2008, MNRAS, 387, 987
\bibitem[]{YL02} Yan, H. \& Lazarian, A. 2002, Phys. Rev. Lett., 89, 281102

\end{thebibliography}
\end{document}